\documentclass[floatfix,
 reprint,
superscriptaddress,
frontmatterverbose,
 amsmath,amssymb,
 aps,pre
]{revtex4-2}
\usepackage[export]{adjustbox}
\usepackage{xcolor}
\usepackage{graphicx}
\usepackage{dcolumn}
\usepackage{comment}
\usepackage{mathtools}
\usepackage{algorithm}

\usepackage{relsize}
\usepackage{mathptmx}
\graphicspath{{GRAPHS/}}

\newcommand{\eref}[1]{Eq.~(\ref{#1})}%
\newcommand{\fref}[1]{Fig.~\ref{#1}} %
\begin{document}

\title{Diffusion in a wedge geometry: First-Passage Statistics under Stochastic Resetting}

\author{Fazil Najeeb}
\affiliation{Cochin university of science and technology, Kalamassery. P.O, Kochi-682022, India}
\author{Arnab Pal}
\affiliation{The Institute of Mathematical Sciences, CIT Campus, Taramani, Chennai 600113, India}
\affiliation{Homi Bhabha National Institute, Training School Complex, Anushakti Nagar, Mumbai 400094,
India}
\author{V.V. Prasad}
\affiliation{Cochin university of science and technology, Kalamassery. P.O, Kochi-682022, India}

\begin{abstract}
 We study the diffusion process in the presence of stochastic resetting inside a two-dimensional wedge of top angle $\alpha$, bounded by two infinite absorbing edges. In the absence of resetting, the second moment of the first-passage time diverges for $\alpha>\pi/4$ while it remains finite for 
 $\alpha<\pi/4$, resulting in an unbounded or bounded coefficient of variation in the respective angular regimes. Upon introducing stochastic resetting, we analyze the first-passage properties in both cases and identify the geometric configurations in which resetting consistently enhances the rate of absorption or escape through the boundaries. By deriving the expressions for the probability currents and conditional first-passage quantities such as splitting probabilities and conditional mean first-passage times, we demonstrate how resetting can be employed to bias the escape pathway through the favorable boundary. Our theoretical predictions are verified through Langevin-type numerical simulations, showing excellent agreement. 
\end{abstract}
%\date{\today}
\maketitle
\section{Introduction}\label{section1}

Diffusion under confinement has attracted considerable attention due to its relevance across a wide range of fields, including intracellular transport in biological systems \cite{cell-book-lewisalberts,lee2021chromatin,Shigesada2003}, reaction-diffusion systems  \cite{Dagdug2024}, socio-economic modeling \cite{pineda2011diffusing,kunitomo1992pricing}, %finance\cite{HIEBER2012165,krugman1991} 
and applications in computer science \cite{song2019generative}. The geometry and nature of confinement have been shown to give rise to rich physical phenomena such as trapping, spatial segregation \cite{lee2021chromatin}, nonlinear mobility \cite{Riefler_2010}, and various forms of anomalous diffusion \cite{banks2005anomalous,hofling2013anomalous}. 
Among the numerous properties that have been investigated, first-passage characteristics to boundaries or targets \cite{Redner_2001,bray2013persistence,firstpassinbounddomain} have been studied extensively, as they govern the time scales associated with the likely completion or persistence of diffusive processes in such systems.

The characteristics of the first-passage time distribution are strongly influenced by both the nature of the confinement and the underlying dynamics of the system. In one-dimensional confined systems  undergoing standard diffusion, the first-passage statistics exhibit exponential decay, implying finite mean first-passage times \cite{Redner_2001}. In contrast, under semi-confined conditions, the distribution is known to follow a power-law decay at long times~\cite{Polya1921,levy1940certains,Redner_2001}. In two-dimensional domains, semi-infiniteness can be incorporated in multiple ways, offering a rich variety of geometries and behaviors. Moreover, two-dimensional systems provide a compelling balance between non-trivial physical features and analytical tractability, making them suitable for modeling a wide range of real-world systems. Additionally, many $N-$particle problems can be effectively mapped onto two-dimensional bounded domains \cite{FISCHER1984}. 
One of the simpler forms of geometric confinement in two-dimensional domains is realized in a region bounded by two infinitely long lines originating from a common point and separated by a fixed angle—commonly referred to as a wedge domain. This domain is semi-infinite: it is unbounded in the radial direction but confined angularly, thereby breaking the spatial homogeneity of the system. Diffusion within wedge domains has been investigated in various contexts. Notably, A. Sommerfeld examined such a configuration in the study of heat conduction \cite{sommerfeld}.
\begin{figure}[b]
    %\centering
    \includegraphics[width=0.48\textwidth]{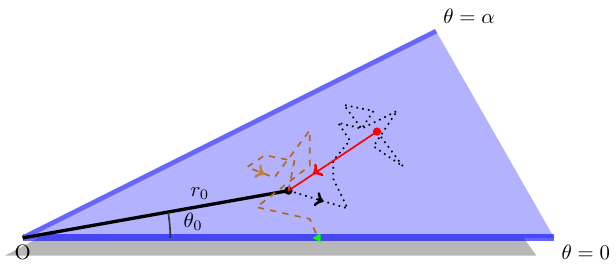}
    \caption{Schematic of a Brownian particle diffusing inside a two dimensional wedge. The particle starts from ($r_0,\theta_0$) and it is being reset intermittently at a rate $\lambda$ to the same location from where it renews the motion. The wedge has two infinite absorbing edges separated at an angle $\alpha$. In this study we analyze the first-passage time to any of these edges (e.g., facilitated by a trajectory in brown dashed) under resetting mechanism.
    }
    \label{f01:brownian-wedge-reset-schematic}
\end{figure}

Beyond serving as a paradigmatic system, the wedge domain model finds relevance in various applied contexts. For instance, in biological systems, a simplified two-dimensional wedge-like geometry has been used to represent the spatial distribution of microtubules within a human cell, aiding in the modeling of virus trafficking through the intracellular medium \cite{virus-effective-motion}. In the realm of reaction-diffusion systems, a general mapping exists where an $N$-particle system on a line -- initialized with the ordering \(x_1 < x_2 < x_3 <...<x_N\) -- can be transformed into a single-particle diffusion process in a conical region in \(R^N\), bounded by the same set of coordinate constraints. Specifically, for $N=3$, this mapping reduces to a single particle diffusing within a two-dimensional wedge-shaped domain with absorbing boundaries \cite{3particletowedgemapping,FISCHER1984,Redner_2001,Krapivsky-Naim_2010,arcsinelaw}.

It has been observed that for diffusive motion within wedge domains, the first-passage time distribution exhibits a power-law decay of the form $ \sim t^{-\beta}$, where the exponent 
$\beta$ depends on the wedge angle \cite{Redner_2001}. Analytical results for such first-passage statistics were initially derived for wedge geometries that are integer subdivisions of the half-plane, i.e., wedge angles of the form $\alpha=\pi/n$, when $n=2$ or  $n$ is odd~\cite{dy}, and when $n$ is any positive integer~\cite{dy2}. This analysis was later extended to arbitrary wedge angles $0<\alpha<2\pi$, by Chupeau et al. \cite{majumder}, who provided a general expression for the survival probability of a diffusive particle within the wedge at all times. Furthermore, the last-passage time statistics—describing the final time a particle arrives at the wedge boundaries—were also determined for two-dimensional wedges, taking the form of a sum of arcsine laws by Comptet and Debois  \cite{arcsinelaw}. More recently, attention has turned to first-passage phenomena in modified wedge environments, such as a Brownian particle subjected to a radial fluid flow within a 2D wedge, which introduces an additional layer of complexity and has become a subject of interest \cite{krapviskyApexflow}.

A natural yet fundamental study in the  first-passage problems would be to design strategies that can control the motion in a systematic manner so that the process lives longer or shorter in the domain of interest \cite{Redner_2001,bray2013persistence}. 
In applications such as target search problems, prolonged survival within the domain may be less desirable than rapid absorption by a target located along the boundary. In transport, similar scenario is observed as faster transit times are more desirable. In ecology, a constant effort is made by the foragers to find the resources in a reasonable timescale.  Thus, a shorter first-passage time is more favorable in most natural settings, as it corresponds to a faster target acquisition.

Towards optimization of the first-passage times, the well studied mechanism of stochastic resetting has been proven to be extremely useful \cite{MR-Evans2011,arnabrauveni,experimentreset,besga2020optimal,evans2020stochastic,pal2024random,jolakoski2023first,kumar2023universal,ray2021mitigating,campos2015phase,domazetoski2020stochastic}. Stochastic resetting refers to a class of models in which the otherwise stochastic evolution of a random spatial variable (e.g., position or velocity) is intermittently interrupted and reset to a specific value or confined to a predefined range of values from where it renews the dynamics \cite{MR-Evans2011}. Systems with stochastic resetting has gained considerable attention in recent years due to  myriads of applications in  describing phenomena in physics and other interdisciplinary areas \cite{evans2020stochastic,arnabrauveni,1dreset, bressloff2020target,schumm2021search,pal2024random}. Besides the plethora of interesting non-equilibrium effects amenable to exact analysis \cite{evans2020stochastic,MR-Evans2011,eule2016non,pal2015diffusion,stojkoski2022autocorrelation}, it has been shown that resetting renders useful beneficial effects to the first-passage time statistics, in particular, leading to a faster completion of search processes in a variety of setups for both open and bounded systems. Recent experiments using optical traps and stochastic robots have showcased interesting avenues in the field \cite{experimentreset,besga2020optimal,paramanick2024uncovering}.

While stochastic resetting has been extensively studied, especially in one dimension, its applications to 2D confined geometries remain less explored ~\cite{bressloff2021asymptotic,ahmad2023comparing,chen2022first,arnab_channel}. The interplay between stochastic resetting and diffusion under geometric constraints offers valuable insights into the efficiency of exploration and the optimization of diffusive processes within confined environments  \cite{activebrwonianmotion-stationary,pal2024channel,arnab_channel,bressloff2021asymptotic,1dreset,chen2022first,ahmad2023comparing,Pal-VVP-PRR2019}. Moreover, the inherently semi-confined nature of the wedge domain naturally motivates an investigation of diffusion dynamics under stochastic resetting. In this work, we explore the influence of stochastic resetting on Brownian motion within wedge domains, focusing on how  resetting strategies impact particle dynamics, particularly in relation to first-passage behavior and transport properties in such spatially constrained geometries. Through a combination of theoretical analysis and numerical simulations, we aim to uncover the mechanisms by which stochastic resetting modifies or enhances first-passage characteristics within wedge-like domains.

The paper is organized as follows: In Sec.~\ref{section2}, we describe the setup of 2D wedge domain and the associated diffusion dynamics within the domain. In the section, we further discuss various statistical quantities, relevant for analysis. 
In Sec.~\ref{section3}, we incorporate stochastic resetting into the framework through the renewal formalism and look at how  the first-passage characteristics is affected. In Sec.~\ref{section4}, we illustrate the existence of optimal mean first-passage time~{MFPT} as a function of the rate of resetting by analysing the coefficient of variation$(CV)$ and construct the phase diagram indicating the parameters for which such optimal resetting rates are present. In Sec.~\ref{section5} we look at the current densities (Sec.~\ref{sec5a}) across the boundaries for the setup, to derive the conditional quantities such as conditional mean exit times (Sec.~\ref{sec5b}) and splitting probability (Sec.~\ref{sec5c}) under resetting. In Sec.~\ref{sec5d}, we demonstrate the recently developed universal criterion for the optimality of conditioned first-passage exits, for the particular case of two dimensional wedge, validating the same. We conclude the paper in Sec.~\ref{section6}, summarizing the outcomes and possible future works. Detailed derivations and discussions on methods of numerical simulations are provided in the Appendix.

\section{Diffusion inside the wedge domain}\label{section2}
We first describe the setup and review the relevant properties of the underlying process of diffusion inside the wedge domain. The wedge domain is the region enclosed by two straight lines radially extending  to infinity, that are separated by a smaller angle denoted as $\alpha$ called the wedge angle. The straight lines act as absorbing boundaries. Considering the polar coordinate representation for the spatial region, the probability distribution function $P_0(r,\theta,t)$  gives the occupation probability density of a Brownian particle to be at $r\in(0,\infty)$ and  $0\le\theta\le\alpha$  at time \(t\), starting from $(r_0,\theta_0)$ [see \fref{f01:brownian-wedge-reset-schematic}]. Another important quantity is the survival probability, describing the probability for the diffusing particles to  survive inside the wedge domain up to a time $t$ without getting absorbed. These quantities were derived earlier \cite{arcsinelaw,Redner_2001,dy,majumder}, for this setup but in the absence of resetting. In this paper, we incorporate the resetting mechanism to explore its impact on the  statistical properties of the system.

\subsection{Probability distribution function}{\label{subsec2a}}
To obtain the probability distribution function~(PDF) \(P_0(r,\theta,t)\) of a  point particle undergoing diffusive dynamics inside the wedge domain, one needs to solve the diffusion equation for the appropriate initial and boundary conditions. 
The \eref{2ddiffusionequation} can be conveniently represented in polar form
\begin{equation}\label{2ddiffusionequation}
    \frac{\partial P_0}{\partial t} = D\left(\frac{\partial^2 P_0}{\partial r^2} + \frac{1}{r}\frac{\partial P_0}{\partial r} + \frac{1}{r}\frac{\partial^2 P_0}{\partial \theta^2} \right).
\end{equation}
For the diffusing particle with the absorbing boundary along the wedge, the initial and boundary conditions are respectively, \(P_0(r,\theta,0) =\delta(r-r_0)\delta(\theta-\theta_0)/r_0\) and \(P_0(r,\theta = 0,t)= P_0(r,\theta = \alpha,t)=0\). The solution for  \eref{2ddiffusionequation} is given as \cite{Redner_2001, dy,majumder,arcsinelaw,FPinwedgedomain}
\begin{equation}
\begin{split}
    \label{pdfgenerictimedomain}
   P_0(r,\theta,t) =& \frac{\exp\left(-\frac{r^2+r_0^2}{4Dt}\right)}{\alpha D t}\\
    &\times\displaystyle\sum_{m=1}^{\infty}\sin\left(\frac{m\pi\theta_0}{\alpha}\right)\sin\left(\frac{m\pi\theta}{\alpha}\right)
    I_{\frac{m\pi}{\alpha}}\left(\frac{rr_0}{2Dt}\right),
\end{split}
\end{equation}
where $I_\mu(x)$ is the modified Bessel function of the first kind.
In the long time limit, the probability distribution  $P_0(r,\theta,t)$ decays as  $\frac{1}{t^{\pi/\alpha+1}}$ which can be easily 
obtained from the following asymptotic expansion of the Bessel function %In this limit ti.e.,(\(x\rightarrow0\)), 
\begin{equation}\label{besselasymptotic}
  I_\mu(x)%|_{x \to 0}  
  \sim \frac{1}{\Gamma(\mu+1)}\left(\frac{x}{2}\right)^{\mu}~~\text{for}~~~x\ll \sqrt{(\mu+1)}.
\end{equation}
This essentially implies that the particle will eventually be absorbed to one of the boundaries in the long time limit.

Unlike the infinite sum in the case of arbitrary wedge angle \eref{pdfgenerictimedomain}, an expression for $P_0(r,\theta,t)$ in terms of a finite sum can be derived (See \ref{pdf_pibyn}) in the case of diffusion inside the wedge~\cite{dy}. When the wedge angle \(\alpha\) is an  integer division of \(\pi\) (i.e.,\(\alpha_n = \frac{\pi}{n}\), with \(n=1,2,3...\)), the PDF follows the equation
\begin{equation}\label{dypdfexpression1}
\begin{split}
     P_0(r,\theta,t)=\frac{1}{4\pi D t}\displaystyle \sum_{k=0}^{n-1}\left(e^{-\frac{R_k^2(x_1,\theta)}{4 D t}}- e^{-\frac{R_k^2(x_2,\theta)}{4 D t}}\right),
\end{split}
\end{equation}
where the exponent \(R_k^2(x,\theta)\) and \(x_1,x_2\) %and \( R_k^2(x_2,\theta)\) 
are defined as  
\begin{equation}\label{discretepdfimplicitvariable}
\begin{split}
    R_k^2(x,\theta) &=r^2+r_0^2-2 rr_0 \cos\left(2\alpha_n k-\theta+x\right), \\
        x_1&=\theta_0, \\
        x_2&=2\alpha_n-\theta_0.
\end{split} 
\end{equation}

The Laplace transform for the  equation \eref{dypdfexpression1} will be useful in calculating the conditional first-passage statistics for wedge angles being integer divisions of $\pi$ for acute angles.
\subsection{Survival probability}\label{subsec2b}
\label{survival-prob- no-resetting}
To understand the  behavior of the first-passage time statistics, it is useful to consider the survival probability $Q_0(r_0,\theta_0,t)$, which estimates the probability that a particle survives in the allowed region without getting  absorbed up 
to time $t$ given that it had started from the initial configuration \((r_0,\theta_0)\).
The expression for the survival probability for the 2D wedge system for time $t$ can be obtained by integrating \(P_0(r,\theta,t)\) in \eref{pdfgenerictimedomain} over the entire range of the wedge domain:
\begin{equation}{\label{survivaltimespace}}
\begin{split}
    Q_0(r_0,\theta_0,t)=&\int_{0}^{\infty} rdr\int_{0}^{\alpha} d\theta P_0(r,\theta,t) \\
    =& \sqrt{\frac{8\pi z_0}{\alpha^2}} e^{-z_0}\displaystyle\sum_{m=0}^{\infty}\frac{\sin[(2\nu_m+1)\theta_0]}{2\nu_m+1}\\&\times\left[I_{\nu_m}(z_0)+I_{\nu_m +1}(z_0)\right],
\end{split}
\end{equation}
where  \( \nu_m = \frac{(2m+1)\pi}{2\alpha}-\frac{1}{2}\) and \(z_0 =\frac{r_0^2}{8Dt}\). The large time asymptotics of the survival probability can be extracted from the above expression. The coefficients of the harmonics in the series \eqref{survivaltimespace} depend on an implicit decaying function of time~[$I_{\nu_m}(z_0)$], whose decay rate increases with $m$. Hence in the large time limit, the only dominant term in the series will be the lowest mode ($m=0$) \cite{Redner_2001}.  Further using the asymptotic expression \eref{besselasymptotic}, one can neglect the term, $I_{\nu_{m}+1}(z_0)$ over $I_{\nu_{m}}(z_0)$,  %since the ratio $I_{\nu_{m+1}}(z_0)/I_{\nu_{m}}(z_0) \sim z_0$ in
and the large time behavior of \eref{survivaltimespace} can be approximated as
\begin{equation}\label{survivallargetimeasymptotic}
    Q_0(r_0,\theta_0,t) \sim \sqrt{z_0}e^{-z_0}\sin(\pi \theta_0/\alpha)I_{\pi/2\alpha-1/2}(z_0),
\end{equation}
leading to the power law behavior of the survival probability $Q_0(t)\sim t^{-\pi/2\alpha}$ at large times~\cite{Redner_2001}. 

One can obtain the behavior moments of the first-passage time towards the absorbing boundaries, of any order $k$, from the expression for the survival probability. In general, the moments of the first-passage time are given using the integral equation 
\begin{equation}
\label{firstpassage-moments-formal-equation}
    \langle t^{k}_0\rangle = -\int_0^{\infty}dt\hspace{0.2em}t^k dQ_0/dt.
\end{equation}
Using the large tail power law behavior of the survival probability in the above equation~[\eref{firstpassage-moments-formal-equation}], the diverging contributions towards the moments can be calculated which shows that the first and second moments ($k=1$ and $k=2$) diverge for $\alpha\geq \pi/2$ and $\alpha\geq \pi/4 $ respectively.
%###############
\section{diffusion under resetting inside wedge}\label{section3}
We now turn our attention to the diffusion dynamics interrupted stochastically by resetting. 
We assume that the underlying dynamics is stochastically interrupted at a constant rate $\lambda$ and the particle is instantaneously reset to a specific location inside the wedge which in this case is taken to be the originating location of the underlying process,  (\(r_0,\theta_0\)), as depicted in \fref{f01:brownian-wedge-reset-schematic}.
To analyze the effect of resetting, we make use of the renewal formalism which essentially allows us to derive the statistical metrics under resetting dynamics from the direct knowledge of the same for the underlying reset-free process.
Following \cite{timedependentreset,1dreset,pal2024random}, we can write
\begin{align}\label{pdfmasterequation}
  P_\lambda(r,\theta,t) = e^{-\lambda t} P_0(r,\theta,t) +  \lambda\displaystyle\int_{0}^{t}d\tau e^{-\lambda\tau} P_0(r,\theta,\tau)Q_\lambda(r_0,\theta_0,t-\tau) .
\end{align}
By performing a Laplace transform in Eq. \eqref{pdfmasterequation} and rearranging the terms, we arrive at
\begin{equation}\label{pdfwithreset}
    \tilde{p}_{\lambda} (r,\theta,s) = \frac{\tilde{p}_0(r,\theta,s+\lambda)}{1-\lambda \tilde{q_0} (r_0,\theta_0,s+\lambda)},
\end{equation}
where \(\tilde{p}_0(r,\theta,s),\hspace{0.5em} \tilde{p}_\lambda(r,\theta,s)\) and \(\tilde{q}_0 (r_0,\theta_0,s)\) are the Laplace transforms of \(P_0(r,\theta,t), \hspace{0.3em}P_\lambda(r,\theta,t),\) and \(Q_0(r_0,\theta_0,t)\) respectively. 
%\(Q_\lambda(r_0,\theta_0,t)\) is obtained using the renewal equation \eref{survivalrenewal}.
Evaluating  the limiting behavior, \(\lim_{s\rightarrow0} s\tilde{p}_\lambda (r,\theta,s)\)  using \eqref{pdfwithreset} one can show that that no steady state exists for any finite value of resetting rate \(\lambda\).

\subsection{Unconditional first-passage dynamics under stochastic resetting}\label{sec3a}
 We look at the characteristics of the first-passage statistics of a diffusive particle subjected to absorbing boundary conditions of the wedge domain. We would specifically like to address how, for the semi-confined domain, resetting alters the mean completion time and when it expedites the process.

We proceed by evaluating the survival probability in the presence of resetting \(Q_\lambda(r_0,\theta_0,t)\). The expression for \(Q_\lambda(r_0,\theta_0,t)\) can be obtained from  the following renewal equation
\begin{equation}\label{survivalrenewal}
\begin{split}
   Q_\lambda(r_0,\theta_0,t) = e^{-\lambda t} Q_0(r_0,\theta_0,t) + \\ \lambda\displaystyle\int_{0}^{t}d\tau e^{-\lambda\tau} Q_0(r_0,\theta_0,\tau)Q_\lambda(r_0,\theta_0,t-\tau).
\end{split}
\end{equation}
Obtaining a Laplace transform on either side of this renewal equation results in
\begin{equation}{\label{survivalwithreset}}
    \tilde{q}_\lambda (r_0,\theta_0,s) = \frac{\tilde{q}_0 (r_0,\theta_0,s+\lambda)}{1-\lambda \tilde{q}_0 (r_0,\theta_0,s+\lambda)}.
\end{equation}
 An expression involving finite sum for the survival probability \(Q_0(r_0,\theta_0,t)\)
has been derived in \cite{majumder}
equivalent to  \eqref{survivaltimespace}  
which upon taking Laplace transform (See \ref{survival_laplace}) renders the following expression:
\par
\begin{figure}[ht]
        \includegraphics[width=1 \linewidth]{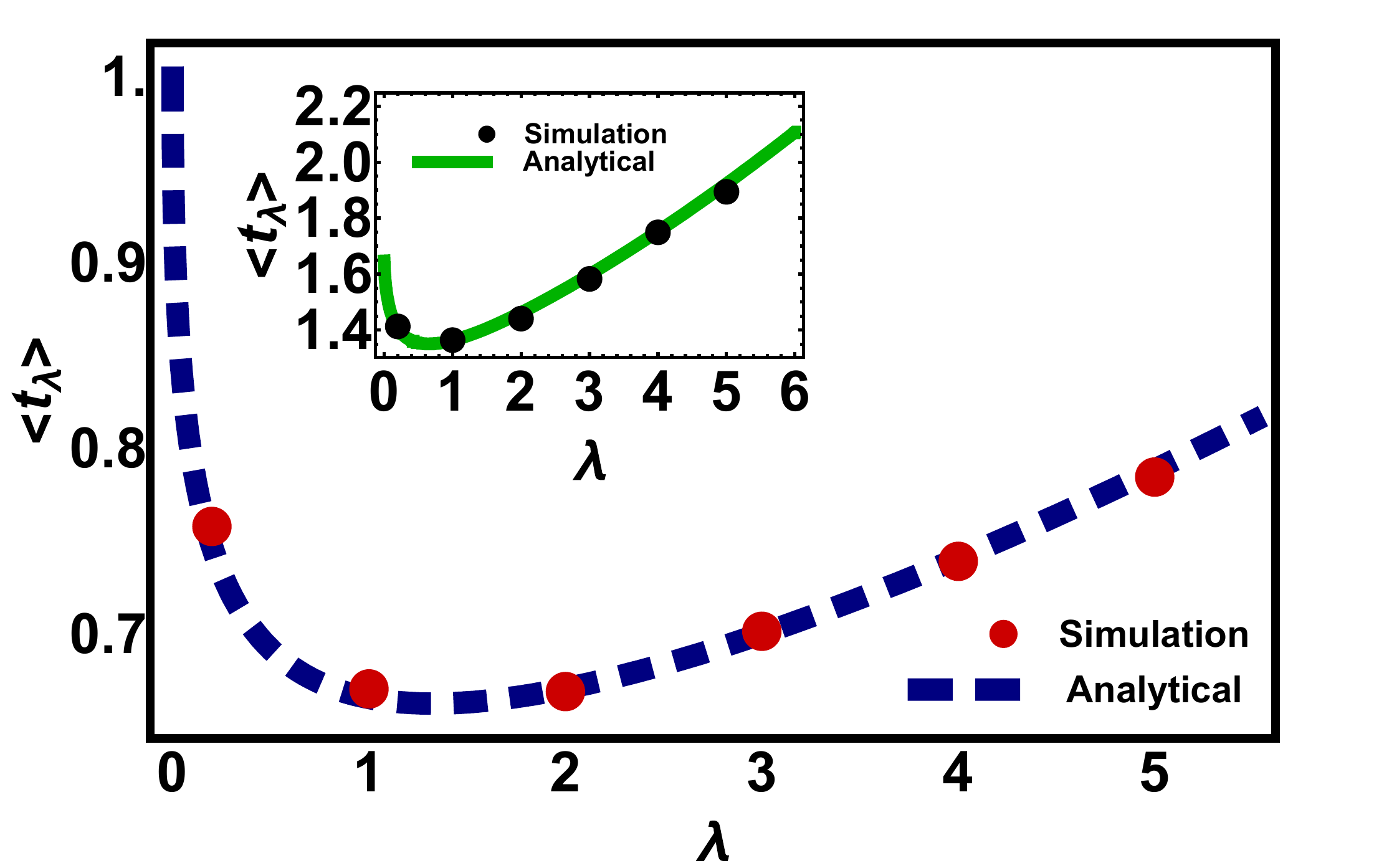}
    \caption{Mean first-passage time $\left< t_\lambda \right >$ vs reset rate $\lambda$ showing optimal behaviour.  The analytical expression for the MFPT [\eref{mfptwithreset}] for two sets of parameters, dashed blue line~(\(r_0 = 2,~\alpha=\pi/3,~\theta_0=\pi/6,~D=1\)) and solid light green line~(inset)~(\(r_0 = 3,~\alpha=\pi/4,~\theta_0=\pi/6,~D=0.5\)) are compared against numerical simulations (Black and red dots respectively), showing good agreement. 
    }
    \label{MFPTvsRESET}
\end{figure}

\begin{widetext}
\begin{equation}\label{survivallaplacedefsum}
    \begin{split}
            \tilde{q}_0(r_0,\theta_0,s) &=\displaystyle \int_{0}^{\infty} dt e^{-st}Q_0(r_0,\theta_0,t) \\
            &=\frac{1-e^{-r_0 \sqrt{\frac{s}{D}}\sin(\text{Min}(\theta_0,\frac{\pi}{2}))}}{s} +\displaystyle\sum _{j=1}^{\frac{\pi }{2 \alpha }+\frac{1}{2}}  \frac{(-1)^j}{s}\left(e^{-r_0 \sqrt{\frac{s}{D}}\sin(\text{Min}(\alpha j -\theta_0,\frac{\pi}{2}))}-e^{-r_0 \sqrt{\frac{s}{D}}\sin(\text{Min}(\alpha j +\theta_0,\frac{\pi}{2}))}\right) \\ &+\frac{r_0}{2\pi \sqrt{s D}}\displaystyle \int_{0}^{\infty}du \left[e^{-r_0\sqrt{\frac{s}{D}}\cosh(\frac{u}{2})}\sinh(\frac{u}{2}) \left(\tan^{-1}\left(\frac{\sin\left(\frac{\pi}{\alpha}(\theta_0 +\frac{\pi}{2})\right)}{\sinh\left(\frac{\pi u}{2\alpha}\right)}\right)+ \tan^{-1}\left(\frac{\sin\left(\frac{\pi}{\alpha}(\theta_0 -\frac{\pi}{2})\right)}{\sinh\left(\frac{\pi u}{2\alpha}\right)}\right)\right)\right].
    \end{split}
\end{equation}
\end{widetext}
The mean first-passage time under resetting can now be obtained as 
\begin{equation}{\label{mfptwithreset}}
     \left< t_\lambda \right > = \lim_{s\rightarrow 0} \tilde{q}_\lambda (r_0,\theta_0,s) =\frac{\tilde{q}_0 (r_0,\theta_0,\lambda)}{1-\lambda \tilde{q}_0 (r_0,\theta_0,\lambda)}.
\end{equation}
In \fref{MFPTvsRESET}, we plot the expression obtained from \eref{mfptwithreset} as a function of reset rate along with the numerical simulations for a particular wedge angle, showing excellent agreement. One finds that for the parameters the mean first-passage time is optimized for a non-zero reset rate in this particular case, indicating that resetting mechanism can be beneficial in resulting in a faster completion of the stochastic process.

\section{CV criterion: Usefulness of the resetting}\label{section4}

It has been shown for diffusing systems that the introduction of resetting minimizes the mean first-passage time for different processes such as absorption to boundaries or searching of a target~\cite{evans2020stochastic}. Probing the question in the context of diffusing particle in a wedge is worthwhile.  The nontrivial feature which can be noticed in the case of the wedge domain is its semi-infiniteness, with the  radial coordinate $r$  unbounded with the angle of the wedge kept fixed. For the geometry, interesting effects on the mean first-passage time exists  even in the absence of resetting~(See Sec.\ref{subsec2b}), due to the long tail behavior of the first-passage times for different wedge angles. A systematic way to check, how resetting dynamics influences the long tailed behaviors of the first-passage process in these distinct set of angles and thereby the mean first-passage times, is by looking at the statistical measure of the coefficient of variation ($CV$) for the underlying process
 \begin{equation}\label{cvdef}
    CV = \displaystyle\sqrt{\frac{ \left<t_{0}^{2}\right>-\langle t_0 \rangle^2}{ \left<t_{0}\right>^2}},
\end{equation}
 in the limit of vanishing resetting rate, i.e., \(\lambda \rightarrow 0 \). Following \cite{arnabrauveni,ahmad2019first}, it can be shown that $CV>1$ is a sufficient criterion for resetting to be useful and furthermore there can exist an optimal resetting rate for which the MFPT is minimum. This criterion was later analyzed in much detail using Landau like expansion in \cite{Pal-VVP-PRR2019} and inspection paradox \cite{inspectionparadox-cvcriterion} outlining the possible physical situations that can result in $CV>1$ \cite{pal2024random}.

\subsection{A wedge phase diagram}
 Following upon the preceding analysis, we note that $CV$ diverges beyond wedge angle $\alpha\geq\pi/4$, where the first moment $\langle t_0 \rangle$ is still finite up to $\pi/2$ but not the second moment $\langle t_0 ^2\rangle$. In this region, the resetting will definitely expedite the process faster than the underlying process.  Non-trivial features emerge for the case with the wedge angle $\alpha<\pi/4$ where both first and second moments are finite and hence $CV$ remains bounded. The variation of $CV$ in this region can be determined using the exact expressions of $\langle t_0 \rangle$ and $\langle t^2_0 \rangle$ which in turn can be obtained respectively by taking the limit $s\rightarrow0$ on the Laplace transform of the survival probability $\tilde{q}_0(r_0,\theta_0,s)$~[\eref{survivaltimespace}] (See \ref{laplace_surv}) and its first derivative $-2 \partial \tilde{q}_0(r_0,\theta_0,s)/\partial s$.
The exercise results in the following expression for first moment of the first-passage time distribution
\begin{equation}
\label{firstmomentzeroreset}
%    \left<t_0\right> = \frac{4r_0 ^2 \alpha^2}{\pi D}\displaystyle \sum_{m=0}^{\infty}\frac{\sin\left(\frac{(2m+1)\pi \theta_0}{\alpha}\right)}{(2m+1)((2m+1)^2\pi^2 -4\alpha^2)},
  \left<t_0\right> = \frac{4r_0 ^2 \alpha^2}{\pi D}\displaystyle \sum_{m=0}^{\infty}\frac{\sin\left([2m+1]\pi \Theta_0\right)}{(2m+1)([2m+1]^2\pi^2 -4\alpha^2)}.
\end{equation}
where,  $\Theta_0\equiv \frac {\theta_ {0}} {\alpha}$. A detailed analysis (See\;\ref{firmomclosed}) of the  series in~\eref{firstmomentzeroreset} yields a compact form for $\langle t_0\rangle$  as shown below
\begin{equation}\label{firstmomentclosedintext}
    \langle t_0 \rangle=\frac{r_ {0}^2}{4D}\left(\frac {\cos\left(\alpha[1-2\Theta_0] \right)}{\cos(\alpha)} - 1 \right),
\end{equation}
which is equivalent to a similar form obtained in \cite{Redner_2001}.
%\frac{r_0^2}{2D} \sin (\theta_0 )\Big(\tan (\alpha ) \cos (\theta_0 )-\sin (\theta_0 )\Big)
%where $\Theta= \frac {\theta_ {0}} {\alpha}$
A series expression could also be obtained  for the second moment
\begin{equation}\label{secondmomentzeroreset}
    \begin{split}
        \left<t_{0}^{2}\right> = &\frac{8 r_0^4 \alpha^4}{\pi D^2}\displaystyle \sum_{m=0}^{\infty}\frac{\sin[(2m+1)\pi \Theta_0]}{(2m+1)} \times \\ &\frac{1}{((2m+1)^2\pi^2 -16\alpha^2)((2m+1)^2\pi^2 -4\alpha^2)}
    \end{split}
\end{equation}
which further leads to a closed form (See \;\ref{secmomentclosed} for the detailed calculation)
\begin{equation}\label{secmomentclosedintext}
\begin{split}
  %  \big\langle t_0^2\big\rangle
%= \frac{r_0^4}{96 D^2}\left(\frac{\cos \left(2 \alpha u_- \right)}{\cos (2 \alpha )}-\frac{4 \cos \left(\alpha u_-\right)}{\cos (\alpha )}+3\right)
\big\langle t_0^2\big\rangle
= \frac{r_0^4}{96 D^2}\left(\frac{\cos \left(2 \alpha[1-2\Theta_0] \right)}{\cos (2 \alpha )}-\frac{4 \cos \left(\alpha[1-2\Theta_0]\right)}{\cos (\alpha )}+3\right).
\end{split}
\end{equation}
%\frac{r_0^4}{96 D^2}\Big(\frac{\cos (2 (\alpha -2 \theta_0 ))}{\cos (2 \alpha )} -8 \sin (\theta_0 )\frac {\sin (\alpha -\theta_0 )}{ \cos (\alpha )}-1\Big)
Using these expressions [Eqs.~\eqref{firstmomentclosedintext} and \eqref{secmomentclosedintext}], we find an exact formula for the coefficient of variation
%\footnotesize
\small
\begin{equation}\label{cvclosed}
     CV=\sqrt{
     \frac{2 (1-5 \cos (2 \alpha )) \cos (\alpha[1 -2 \Theta_0])+5 \cos (\alpha )+3 \cos (3 \alpha )}{12 \cos (2 \alpha )\Big(\cos(\alpha[1 -2\Theta_0])- \cos(\alpha )\Big) }}
\end{equation}

Now one could readily observe that the $CV$ is independent of both the initial (and resetting) radial coordinate $r_0$ and the coefficient of diffusion $D$, but only a function of the initial (and resetting) angular coordinate \(\theta_0\) and the wedge angle \(\alpha\) through the function $\Theta_0(=\theta_0/\alpha)$. Thus the characteristics of whether or not the resetting behavior is advantageous in making the mean first-passage optimal, can be visualized by evaluating $CV$ on a \(\theta_0-\alpha\) parameter space with the domain  limited by $0<\theta_0<\alpha$.% as displayed in \fref{cvphasediagram}.

A close inspection of the %series(removed since we have closed forms) 
expressions for the moments-either in terms of the series [Eqs.~\eqref{firstmomentzeroreset},~%,secondmomentzeroreset,firstmomentclosedintext,secmomentclosedintext}), 
and \eqref{secondmomentzeroreset}] or the closed forms [Eqs.~\eqref{firstmomentclosedintext} and \eqref{secmomentclosedintext}]- also provides the information on the parameter range for which the moments indicate qualitatively different behaviour. One can see from \eref{firstmomentclosedintext} that, due to the term $\cos(\alpha)$ in the denominator, %that
the first moment diverges at \(\alpha=\pi/2\) %due to the term $m=0$ in the denominator,
with no physically valid solution for \(\pi/2<\alpha<2\pi\). Similarly, the second moment  (See \eref{secondmomentzeroreset} or \eqref{secmomentclosedintext})  does not have finite value as \(\alpha\rightarrow\pi/4\), due to the $\cos(2\alpha)$ term in the denominator.
It suffices therefore to restrict the CV analysis within the region \(0 \leq \alpha <\pi/4\) of the wedge domain, where both first and second moments are positive definite.  As noted earlier~[See Sec. \ref{subsec2b}], for the wedge angles \(\alpha>\pi/4\), the moments diverge making resetting beneficial trivially to optimize the mean first-passage time for any choice of resetting position.
 \begin{figure}[ht]
    \includegraphics[width=1 \linewidth]{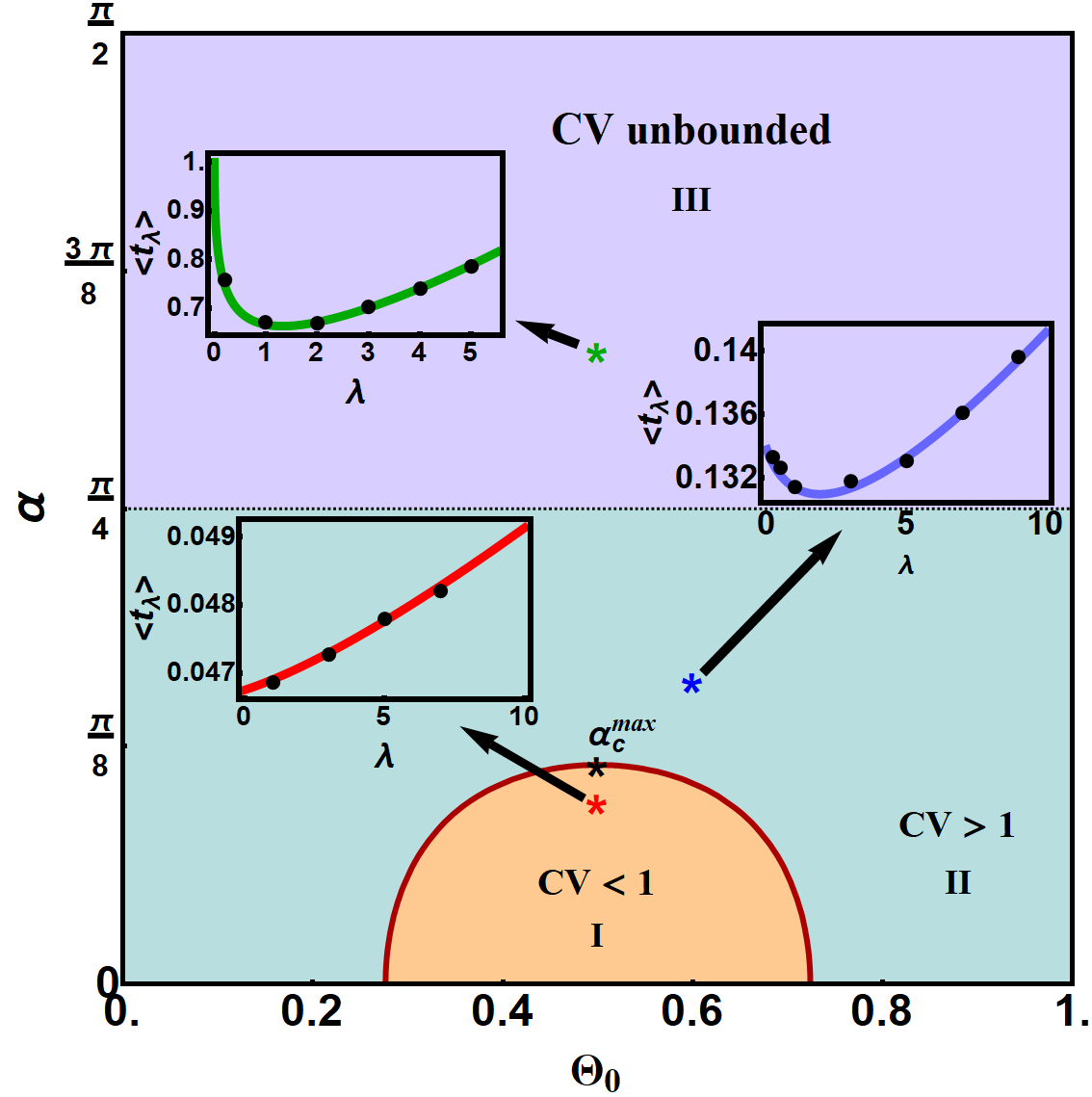}
    %\captionsetup{justification=justified}
    \caption{The coefficient of variation (CV) plotted as a function of system parameters, with the vertical axis representing, wedge angle $\alpha$ and the horizontal axis representing the ratio $\Theta_0\equiv\theta_0/\alpha$ with $\theta_0$ being the initial angle ($0<\theta_0\le\alpha$).
Three distinct regions are identified: In region 
\textbf{I} (depicted in orange)), $\text{CV} < 1$, indicating that resetting does not expedite the process for any $(\theta_0, \alpha)$ %(Red asteric symbol $*$ represents the point $\alpha=0.3, \theta_0/\alpha=0.5$).
and in region \textbf{II} (in cyan),  $\text{CV} > 1$, where resetting can optimize the MFPT.
%(Blue $*$ represents the point $\alpha=0.5, \theta_0/\alpha=0.6$) 
In region \textbf{III} (in purple), the second moment has no finite value while the first moment remains finite, causing the CV to diverge -- resetting is always beneficial in such cases with MFPT attaining a global minimum for $\lambda > 0$. The point(Black asterisk) corresponding to the maximum of the critical wedge angle $\alpha_c^{max}$ has also been represented in the figure. %(Red $*$ represents the point $\alpha=\pi/3, \theta_0/\alpha=0.5$). 
The colored asterisk symbols(Blue, Green and Red) in each region represents parameter values for which  MFPT is plotted as a function of $\lambda$ (in the inset) corroborating the above findings.%{\color{red}NOTE**The above figure has been updated using the newly derived closed form expression for $CV$. Also the critical wedge angle has been included with other subtle changes}
}
    \label{cvphasediagram}
\end{figure}

 Fig.(\ref{cvphasediagram}) displays a phase diagram with domains for the relevant range of wedge angle $\alpha$ %(as per discussion above) 
and the fraction of resetting angle to the wedge angle \(\Theta_0(\equiv\theta_0/\alpha)\) with $0<\theta_0\le\alpha$, where resetting can be beneficial to expedite the mean first-passage time  or otherwise, based on the CV analysis. In the plot, the middle region at the bottom  depicted in light orange color (color online) describe the region where $CV<1$, ie., where resetting is not beneficial. The region in the middle, depicted in cyan (color online) is where $CV>1$ where the  resetting will be beneficial to optimize the mean time. The pattern illustrate that for small values of the wedge angle $\alpha$, there exist initial angles where diffusive dynamics is the dominant process to result in accessing the boundary faster and to make MPFT minimal. However for higher $\alpha$ this turns out not to be the case, where diffusive dynamics result in higher frequency of large range excursions to the unbounded radial direction making resetting the dominant contribution to lower the mean first-passage time. A notable feature that can be realized from the diagram is the possibility that even in the infinitesimal values of wedge angle, there are initial conditions where the diffusive mechanism alone does not lead to an
optimal mean completion time of the process. The symmetry in the plot about $\Theta_0=1/2$ %$\theta_0/\alpha=1/2$
line may as well be noticed from the relation, %$CV(\theta_0/\alpha=1/2-\delta)=CV(\theta_0/\alpha=1/2+\delta)$
$CV(\Theta_0=1/2-\delta)=CV(\Theta_0=1/2+\delta)$
for any $0<\delta<1/2$~[See \eref{cvclosed}].
%Eqs.~(\ref{firstmomentzeroreset}) and (\ref{secondmomentzeroreset})]. 

\bigskip

\subsection{Determining critical wedge angle $\alpha_{c}$}
By imposing the condition \(CV=1\) in \eref{cvclosed}, we can define a locus of critical wedge angles \(\alpha_c\) that demarcates the regions \textbf{I} and \textbf{II} in Fig.~\ref{cvphasediagram}. For \(\alpha>\alpha_c\), one has \(CV>1\), where stochastic resetting is always beneficial. Substituting $CV$ to be unity in \eref{cvclosed} and inverting yields the critical value \(\theta_{0c}\) (or, in scaled form, \(\Theta_{0c}\)) as a function of \(\alpha_c\) as follows

\begin{equation}\label{thetaalpharel}
   \Theta_{0c} =\frac{\theta_{0c}}{\alpha_c}=\frac{1}{2}\pm\frac{1}{2\alpha_c } \cos ^{-1}\left(\frac{11 \cos (\alpha_c )+9 \cos (3 \alpha_c )}{2 (11 \cos (2 \alpha_c )-1)}\right).
\end{equation}
The above equation 
%[\eref{thetaalpharel}], evidently %captures 
evidently
follows from the underlying symmetry of $CV$ about the bisecting angle of the wedge denoted by $\Theta_0=1/2$. 

The upper bound of the critical wedge angle $\alpha^{max}_c$, above which resetting is always helpful to optimize the mean first passage time, happens to be at exactly $\Theta_{0c}=1/2$ on the $CV=1$ separatrix curve. Substituting $\Theta_{0c}=1/2$ 
into \eref{thetaalpharel}, we can compute the upper bound $\alpha^{max}_c$ from the following relation
\begin{equation}\label{alphacritical1}
\begin{split}
    \cos ^{-1}\left(\frac{11 \cos (\alpha_{c}^{max} )+9 \cos (3 \alpha_{c}^{max} )}{22 \cos (2 \alpha_{c}^{max} )-2}\right)=0,
\end{split}
\end{equation}
and hence finding the solution for
\begin{equation}\label{alphacritical2}
\begin{split}    
    9\cos (3 \alpha_{c}^{max} )-22 \cos (2 \alpha_{c}^{max} )+11 \cos (\alpha_{c}^{max} )+2=0.
\end{split}
\end{equation}
Equation \eqref{alphacritical2} can be  solved (See \ref{solve-alphac}) to obtain a unique value for $\alpha_{c}^{max}$ within the region of validity ($0<\alpha<\pi/4$) given by
\begin{equation}
    \alpha_{c}^{max}=\cos^{-1}\left(\frac{1+\sqrt{55}}{9}\right)\approx\frac{\pi}{8.67}\; \text{rad}.
\end{equation}

\section{conditional first-passage dynamics under stochastic resetting} \label{section5}
A relevant observable in the context of diffusion within the infinite absorbing wedge boundaries is related to the statistics of exit of particle to a specific  boundary and their first-passage times. Such statistics are important in several scenarios as in diffusion reaction systems where the reactant is in the vicinity of multiple substrates with one of them being more favorable compared to others. A pertinent question then would be to ask the statistics of accessing the favorable target or the typical time to access that. Similar considerations arise in biological systems as well, where the survival of a bacteria  \textit{E.Coli}, stuck between a bacteriophage rich environment and nutrient molecules, may depend on what it encounters first \cite{ecoliphage}. 

In particular, there exists a one-to-one correspondence to three particles diffusing on a line,  to the single-particle diffusion inside the wedge domain with the dynamics modeled to terminate as soon as any pair of the particles crosses paths\cite{arcsinelaw}. Such setups are interesting in the context such as,
a reactant %(the middle particle \textbf{B}) 
encountering  the nearest substrate %(\textbf{A} and \textbf{C}) 
with one of them favorable  when compared to another, %(See \fref{3particle})
and all undergoing Brownian motion~\cite{multisiteenzyme}.
The setup may also be looked at to model dynamics of  a substrate and an inhibitor competing for the same active site on an enzyme resulting in distinct effects of enzymatic activity~\cite{competetive-inhibitors}.

\par Further, the diffusion of a single particle in a wedge domain is of significance on its own, as it addresses the scenario where the interplay between non-trivial geometry and dynamical properties influence the preferential absorption to one of the boundaries and how and when resetting turns out to be beneficial for the process.
In the following section, we analyze these conditional aspects of the first-passage process under stochastic resetting, in detail.

\subsection{Probability flux or current through the wedge boundaries}\label{sec5a}
The conditional first-passage quantities can be obtained once we  enumerate the fraction  of flux contributions through each of its absorbing boundaries. In this section we define the current density under reset in 2D-polar coordinates and determine the exact expression for the same, at the wedge boundaries.
Using \eref{2ddiffusionequation}, we can define a general expression for the 2D probability current density  \(\vec J_\lambda(r,\theta,t)\) through an arbitrary point \((r,\theta)\)  inside the wedge domain. By definition, the probability flux is the negative gradient of the probability distribution function \(P_\lambda (r,\theta,t)\)
\begin{equation}
    \label{definecurrent}
    \vec J_\lambda (r,\theta,t) = - D \vec\nabla_{r,\theta}P_\lambda (r,\theta,t),
\end{equation}
where
\[\vec\nabla_{r,\theta}\equiv \hat{r}\frac{\partial}{\partial r} + \hat{\theta}\frac{1}{r}\frac{\partial}{\partial\theta}. \]
At any point inside the wedge domain or on its boundaries, the flux will be the sum of its two components: The radial flux (\(J_{\lambda}^{r}\)) and the angular flux (\(J_{\lambda}^{\theta}\))
\begin{equation}
    \label{fluxcomponentform}
    \vec J_{\lambda} = J_{\lambda}^{r} \hat{r}+  J_{\lambda}^{\theta} \hat{\theta}.
\end{equation}
Note that we are particularly interested in the current at the boundaries i.e. at $\theta=0, \alpha$. As the contribution of the radial component of the flux is identically zero at the boundaries, \eref{fluxcomponentformboundary} will have only angular contributions and one can define these as 
\begin{equation} \label{fluxcomponentformboundary}
\begin{split}
  J_{\lambda}^{+}(r,t)& =J^\theta_{\lambda}(r,\theta=\alpha,t), \\
   J_{\lambda}^{-}(r,t)& =J^\theta_{\lambda}(r,\theta=0,t).
\end{split}
\end{equation}
Where the superscripts $+$ and $-$ on the left hand side of \eref{fluxcomponentformboundary} indicates  the currents $J^{\pm}_\lambda(r,t)$ through the wedge boundaries with $\theta=\alpha$ and $\theta=0$ respectively. The conditional quantities such as the conditional mean exit times and splitting probabilities, through the wedge boundaries, can be calculated by enumerating the fraction of exits through each boundaries over all time. These time integrated quantities can be conveniently be obtained using the Laplace transform of currents
$\int_0^\infty e^{-st}J^\pm_\lambda(r,t)dt=\tilde{j}^\pm_{\lambda}(r,s), $ %\cite{1dreset}, 
where the former %time integration of current 
can be evaluated by taking the limit $\displaystyle\int_0^{\infty}J^{\pm}_\lambda(r,t)dt=\displaystyle\lim_{s\rightarrow0}\tilde{j}^\pm_{\lambda}(r,s)$.
In the following, we consider the conditional first-passage quantities for the special case where the wedge angles  are integer divisions of $\pi$,~i.e., $\alpha_n =\pi/n$,~(with $n=1,2,3...$), for which the probability distribution function is given in \eref{dypdfexpression1} in terms of a finite sum. Besides providing compact results, this scenario provides deeper insights to analyze and understand the essence of the problem.

To obtain the expression for the fluxes  $\tilde{j}_{\lambda}^{\pm}(r,s)$  through the boundaries, one proceeds as follows. By taking the Laplace transform of \eref{definecurrent}

\begin{equation}\label{defineboundarycurrent}
\begin{split}
    \tilde{j}_{\lambda}^{+}(r,s)= -\frac{D}{r}\frac{\partial \tilde{p}_\lambda (r,\theta,s)}{\partial \theta}|_{\theta=\alpha}, \\ \tilde{j}_{\lambda}^{-}(r,s)= +\frac{D}{r}\frac{\partial \tilde{p}_\lambda (r,\theta,s)}{\partial \theta}|_{\theta=0},
\end{split}
\end{equation}
and substituting $\tilde{p}_\lambda(r,\theta,s)$ evaluated using 
the Laplace transform of \eref{dypdfexpression1}
\begin{equation}\label{pdflaplacedefsum}
\begin{split}
     \tilde{p}_0(r,\theta,s) &= \displaystyle \int_{0}^{\infty} dt e^{-st}P_0(r,\theta,t) \\ &=\frac{1}{2\pi D} \displaystyle \sum_{k=0}^{n-1}\left[K_0\left(R_k(x_1,\theta)\sqrt{\frac{s}{D}}\right) \right. \\
     &\left.-K_0\left(R_k(x_2,\theta)\sqrt{\frac{s}{D}}\right)\right],
\end{split}
\end{equation}
in \eref{pdfwithreset}, one obtains $\tilde{j}^\pm_{\lambda}(r,s)$,   resulting in expressions
\begin{equation}\label{jplus}
     \tilde{j}_{\lambda}^{+}(r,s) =  \frac{-r_0\eta_s}{\pi[1-\lambda q_0(s)]}\displaystyle \sum_{k=0}^{n-1} \sin\left([2k-1]\alpha_n+x_1\right) \frac{K_1\left(\eta_sR_k(x_1,\alpha_n)\right)}{R_k(x_1,\alpha_n)}.
\end{equation}
\begin{equation}\label{jminus}
    \begin{split}
         \tilde{j}_{\lambda}^{-}(r,s)=  \frac{r_0\eta_s}{2\pi[1-\lambda q_0(s)]}\displaystyle \sum_{k=0}^{n-1}\left[\sin\left(2\alpha_nk+x_1\right) \frac{K_1\left(\eta_sR_k(x_1,0)\right)}{R_k(x_1,0)} - \right.\\ \left. \sin\left(2\alpha_nk+x_2\right)\frac{K_1\left(\eta_sR_k(x_2,0)\right)}{R_k(x_2,0)}\right] .
    \end{split}
\end{equation}
where \(R_k(x_1,0), R_k(x_2,0)\)  and \(R_k(x_1,\alpha_n)\)
%\(R_k(x_2,\alpha)\)
are obtained by substituting values for \(\theta\) in \eref{discretepdfimplicitvariable}, corresponding to the boundary of evaluation. The symbol $\eta_s = \sqrt{(s+\lambda)/D}$.

\begin{figure}[b]
    \includegraphics[width=0.5\textwidth,left]{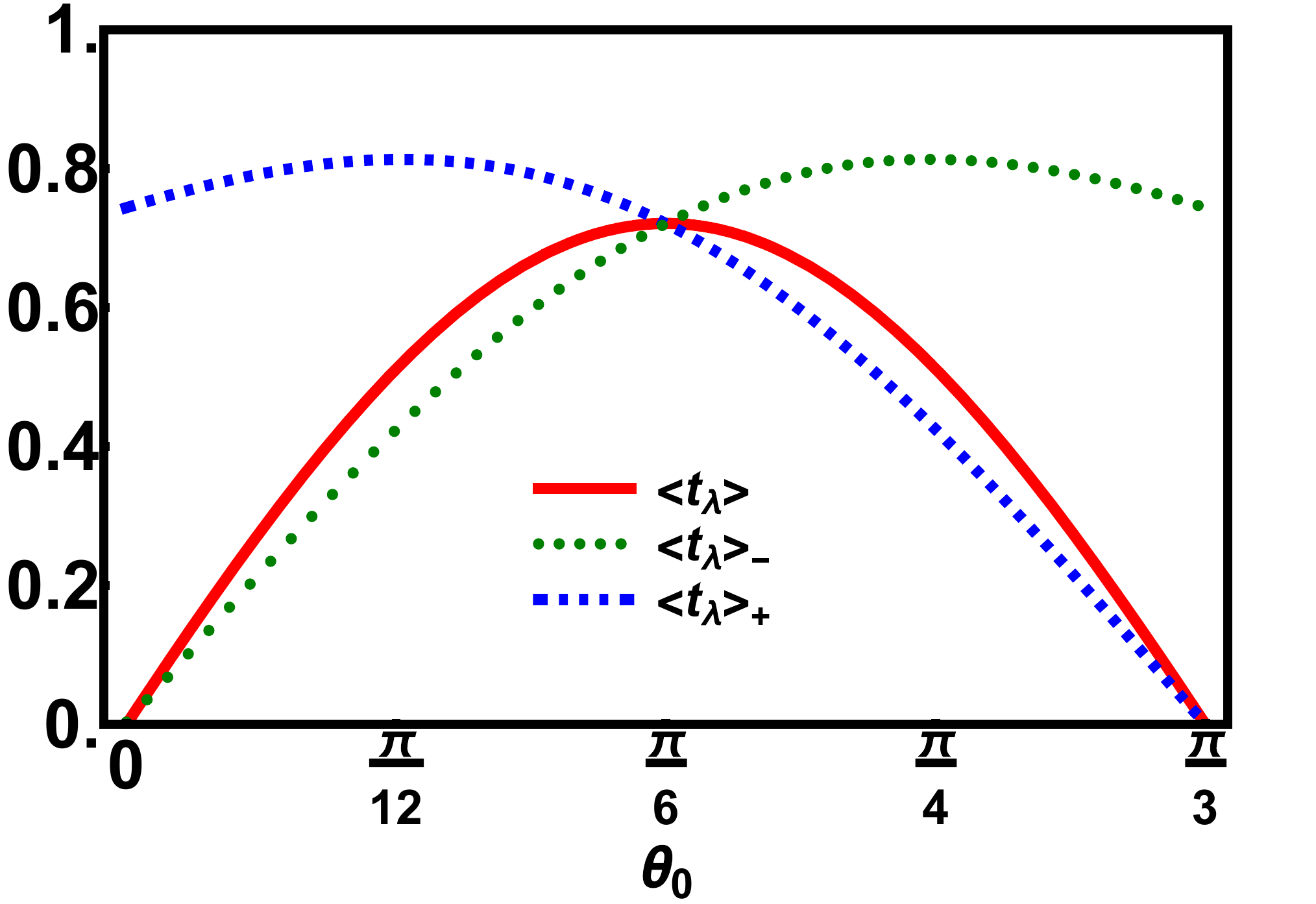}
    \caption{Plot of $\left<t_\lambda\right>_{\pm}$ [\eref{cmfpttimedomain}] and $\left<t_\lambda\right>$ [\eref{mfptwithreset}] plotted as a function of the reset angle $\theta_0$, evaluated for the parameters, $\alpha=\frac{\pi}{3},~\lambda=0.3,~D=1~\text{and}~r_0=2.$ 
    }
    \label{cmfpt-mfpt}
\end{figure}
\subsection{Conditional mean first-passage times}\label{sec5b}
\begin{figure}[t]
\centering
\includegraphics[width=0.5\textwidth]{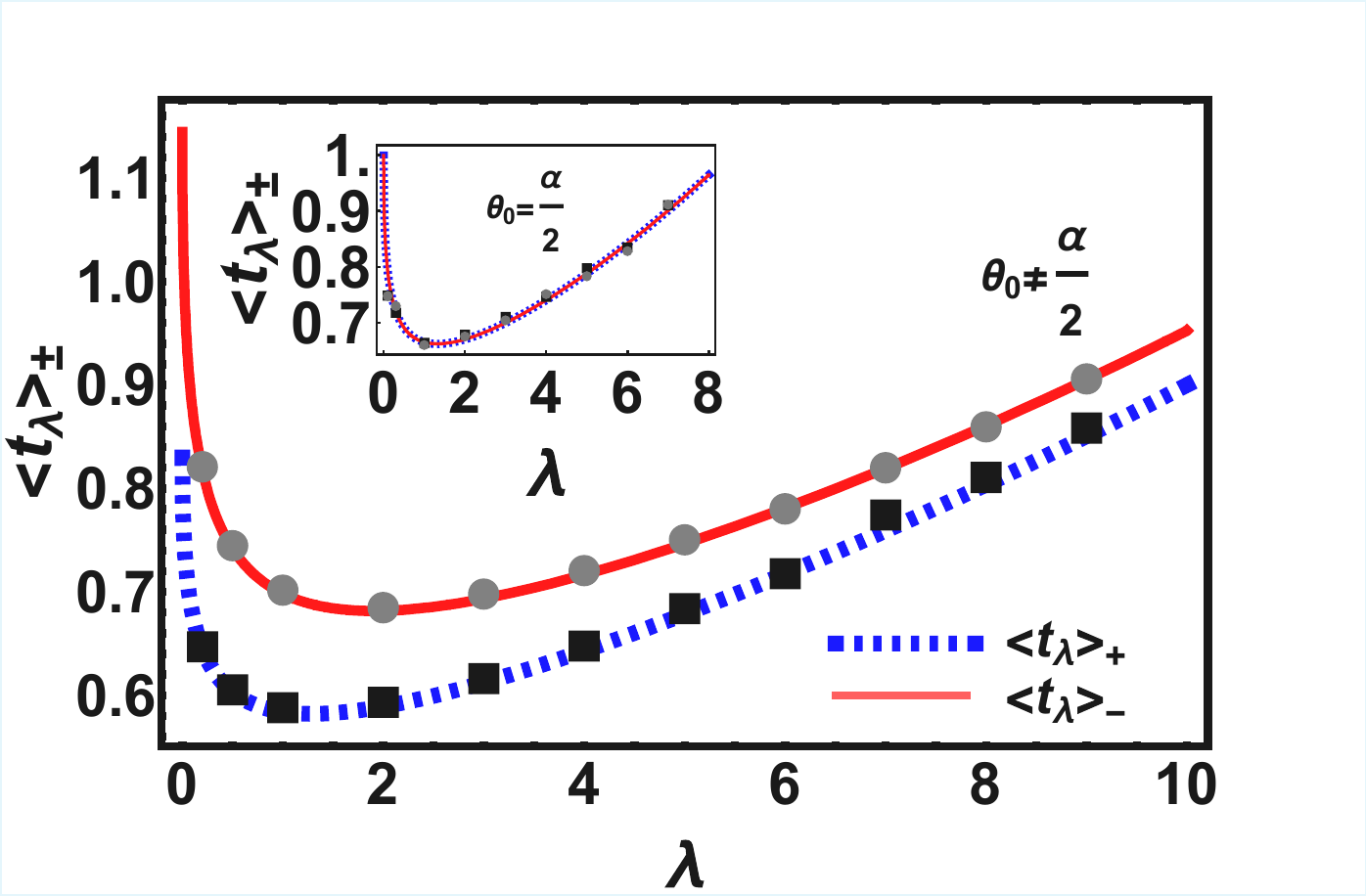}
\caption{ Plot showing conditional mean first-passage times (CMFPT) versus reset rate $\lambda$, for both asymmetric ($\theta_0 \ne \alpha/2$) and for symmetric $\theta_0 = \alpha/2$~(inset) resetting from the analytical expression given in ~\eref{cmfpttimedomain}. The parameters are, $\alpha = \pi/3,~r_0 =2,~D=1$, with $\theta_0=\pi/5$~(main plot) and $\theta_0=\pi/6$~(inset). CMFPT is seen to attain an optimal value for the parameter values.  Black squares and gray dots respectively in the main and inset plots are data points obtained from numerical simulations. 
}
\label{cmfptplots}
\end{figure}

To determine the conditional mean first-passage times through each of the boundaries, \(\left<t_\lambda\right>_{\pm}\), we calculate the total current through each boundary.  The probability current density represents the flux of particles across a boundary per unit radial length. Thus to enumerate the total fraction of particles escaping through the whole boundary, we have to integrate \(J_{\lambda}^{\pm}(r,t)\) over the entire boundary over all time. Now the conditional mean first-passage time through a given boundary is related to the net flux by the following relation~\cite{Redner_2001,1dreset}:
\begin{equation}\label{cmfpttimedomain}
   \left<t_\lambda\right>_{\pm} = \frac{\displaystyle\int_{0}^{\infty} dr\displaystyle \int_{0}^{\infty} dt\hspace{1pt}t J_{\lambda}^{\pm}(r,t)}{\displaystyle\int_{0}^{\infty} dr\displaystyle\int_{0}^{\infty}J_{\lambda}^{\pm}(r,t)dt}= \frac{-\displaystyle\int_{0}^{\infty} dr \frac{\partial \tilde{j}^{\pm}_{\lambda}(r,s)}{\partial s}|_{s\rightarrow 0}}{\displaystyle\int_{0}^{\infty} dr \tilde{j}^{\pm}_{\lambda}(r,s=0)},
\end{equation}
where, \(J_{\lambda}^{\pm}(r,t)\) are the currents evaluated at the boundaries in time domain and $\tilde{j}_{\lambda}^\pm(r,s)$ are their  respective Laplace transforms.
We avoid the long expressions for brevity and present the plots (\fref{cmfptplots}) to show the behavior of the conditional mean first-passage times with respect to the different parameters.
\subsection{Splitting probabilities}\label{sec5c}
\begin{figure}[t]
    \centering
    \includegraphics[width=0.5\textwidth]{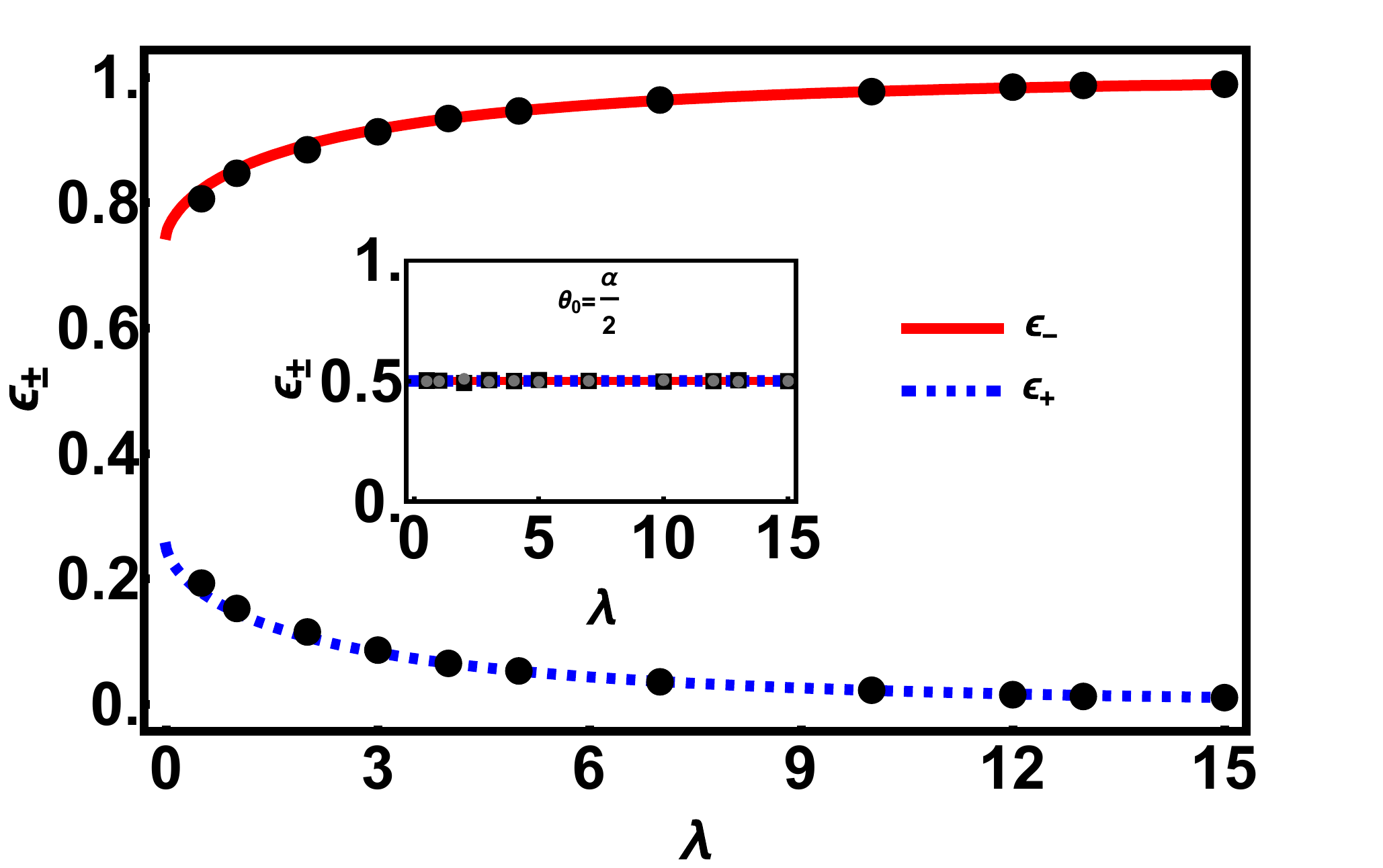}
    \caption{Splitting probability  $\epsilon_\pm$ plotted against reset rate $\lambda$ using \eref{splittingprobdeftime}. The splitting probabilities $\epsilon_+$ and $\epsilon_-$ are seen to approach 0 and 1 respectively for higher reset rates, given the resetting is done asymmetrically in the wedge domain with the following parameters, $\alpha = \pi/2,~r_0 =2,~\theta_0=\pi/8,~D=1$. The plot given as inset is for symmetric resetting with $\theta_0=\pi/4$ with the other parameters kept same. The black dots in the main plot (and gray dots and black squares for inset) represents numerical simulations done for the corresponding parameters validating the analytical expressions.}
    \label{fig:splitting probability}
\end{figure}

Another important set of conditional statistics that one could evaluate for the system is related to the splitting probability $\varepsilon$.  The splitting probability provides the probability that a stochastic process initialised at \(\theta = \theta_0\) and \(r=r_0\) inside the wedge domain, will eventually get absorbed by one of the boundaries without terminating in the other. As in the case of one dimensional systems, one finds that, when resetting dynamics is added, the absorption through one of the boundary could be chosen  over the other, even for a slightest asymmetry introduced to the initial conditions, with respect to the wedge geometry i.e., \(\theta_0 \not=\alpha/2\). Despite the stochasticity in movement towards the boundaries, the preferential exit is guaranteed with the splitting probability approaching unity for higher reset rate~[See \fref{fig:splitting probability}].
To evaluate the splitting probability, we need to enumerate the cumulative fraction of processes that end at a given boundary over all time. % \(t \rightarrow \infty\).
This can be obtained by integrating the boundary fluxes, \eref{jplus} and \eref{jminus} over the entire radial range \(0\le r \le \infty\), for all time \(0\le t < \infty\) as follows:
\begin{equation}\label{splittingprobdeftime}
    \epsilon_\pm = \displaystyle\int_0^\infty dr  \displaystyle\int_0^\infty dt J_\lambda^\pm(r,t)=\displaystyle\int_0^\infty dr \tilde{j}_\lambda^\pm(r,s=0),
\end{equation} 
where, \( \epsilon_+\) and \(\epsilon_- \) are splitting probability through the boundaries at \(\theta = \alpha\) and \(\theta = 0\)  respectively.

\subsubsection{Symmetric resetting}
When the process resets along the angular bisector of the wedge, the splitting probability becomes equal. This can be seen by substituting \(\theta\) values corresponding to either boundaries and \(\theta_0 = \pi/{2n}\) for the symmetric initiation in Eqs.~(\ref{jplus}) and (\ref{jminus}). Even from the symmetry of the setup, one can see that for any value of resetting rate the fluxes through the either boundaries are equal and since \(\epsilon_+ + \epsilon_- = 1\), the splitting probabilities \(\epsilon_{\pm} = 1/2\).
\subsubsection{Asymmetric resetting}
When the process is reset away from the wedge bisector (\(\theta_0\not=\alpha/2\)), frequent resetting increases the bias of exit probability towards the boundary to which it is closer as it is made to reset. This happens at the cost of diminishing exit probability through the farther boundary. With sufficient resetting (\(\lambda \gg 0\)), an asymmetric relocation  will lead to a preferential exit through one of the absorbing boundaries with near definiteness (\(\epsilon \rightarrow1\)) .
\par
One can see that \eref{cmfpttimedomain} can be rearranged to obtain the relation \(\langle t_\lambda\rangle = \epsilon_+ \langle t_\lambda \rangle_+ + \epsilon_-  \langle t_\lambda\rangle_-\) between the conditional and unconditional mean exit times. In light of this relation, the diffusion inside wedge domain with two absorbing boundaries under reset, may be viewed as a Bernoulli-like first-passage process, with a tunable bias \cite{BernoulliTrial,1dreset}. 
\subsection{Analysis on the universal criterion for conditional outcomes under stochastic resetting}\label{sec5d}
\begin{figure}
    \centering
    \includegraphics[width=1\linewidth]{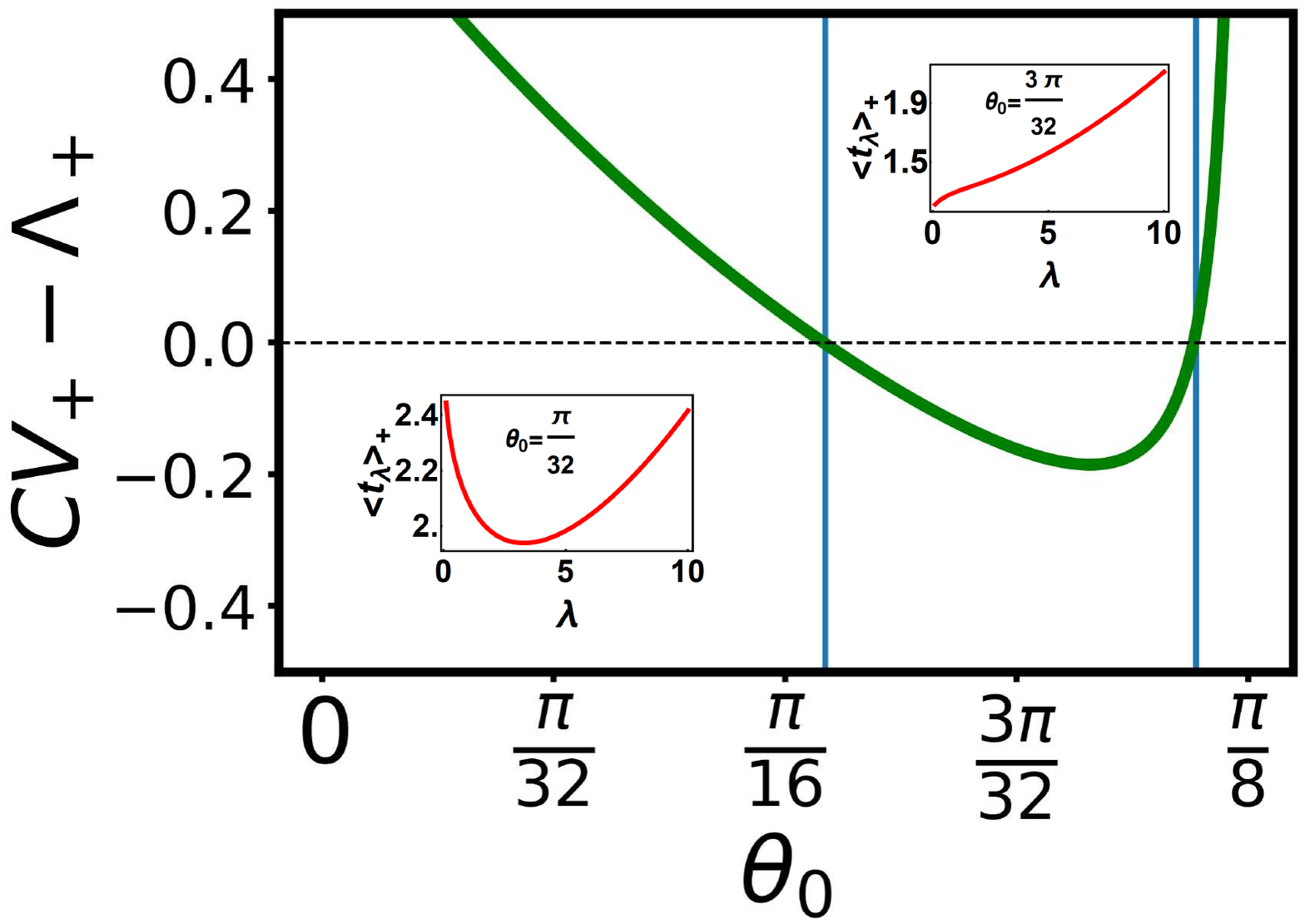}
    \caption{ Variation of $CV_+ -\Lambda_+$ against  the resetting angle $\theta_0$ for a fixed wedge angle $\alpha=\pi/8$ ($D=1,~r_0 =10$) is plotted (solid green line).  Distinct regimes are demarcated by solid blue vertical lines which distinguishes the cases  $CV_+>\Lambda_+$,  wherein  the conditional MFPT ($\langle t_\lambda \rangle_+$ in this case)  is minimized for $\lambda>0$ from the cases  where $CV_+<\Lambda_+$,  with no optimal reset rate for the conditional MFPT. In the inset, conditional MFPT is plotted (as red solid curves) as a function of reset rate $\lambda$ for representative parameters from each regime illustrating the existence or otherwise of the non-zero optimal reset rate.}
    \label{criterioncmfpt}
\end{figure}

In this section we look into the universal criterion, introduced in \cite{cmfptcriterion,pal2025optimal}, which is a sufficient condition to determine whether stochastic resetting can minimize the conditional outcomes such as the conditional mean first-passage times $\langle t_\lambda\rangle_\pm$.  The criteria for minimization of the conditional exit times for $\lambda>0$ which implies the existence of a finite $\lambda$ for which $\langle t_\lambda \rangle_\pm <\langle t_{0}\rangle_\pm$, is found to be \cite{cmfptcriterion} 
\begin{align}
    CV_\pm > \Lambda_\pm,
    \label{Condition for optimality-conditional}
\end{align}  where  
\begin{align}
\label{cvplusminus}
    %CV_\pm=\sqrt{(\langle t_{0}^2\rangle_\pm - \langle t_0 \rangle_\pm^2)/ \langle t_0 \rangle_\pm^2}
    CV_\pm=\sqrt{\frac{\langle t_{0}^2\rangle_\pm - \langle t_0 \rangle_\pm^2}{\langle t_0 \rangle_\pm^2} }
\end{align} is the relative fluctuation of the conditional first-passage times for the underlying resetting free process and 
\begin{comment}
\label{lambdaplusminus}
\Lambda_\pm=\sqrt{(\langle t_0 \rangle^2/2\langle t_0 \rangle_\pm^2)[1+CV^2]}
\end{comment}
\begin{align}
    \Lambda_\pm=\sqrt{\frac{\langle t_0 \rangle^2}{2\langle t_0 \rangle_\pm^2}[1+CV^2]}
\end{align}
is a bound constructed from the first-passage observables for the underlying process. 
To proceed further we compute the mean $\langle t_0\rangle_\pm$ from Eq. (\ref{cmfpttimedomain}) and the second moment $\langle t_0^2\rangle_\pm$ from the following relation (by setting $\lambda = 0$ for the underlying processes)
\begin{equation}
     \langle t_0^2\rangle_\pm=\frac{\displaystyle\int_0^\infty dr \frac{\partial^2\tilde{j}_{\lambda=0}^\pm}{\partial s^2}\big|_{s\rightarrow0}}{\epsilon_\pm(\lambda=0)}.
\end{equation}
To  demonstrate  the condition (\ref{Condition for optimality-conditional}) for the wedge geometry, we have plotted 
$CV_+ -\Lambda_+$ as a function of  $\theta_0$ for a given $\alpha$ in \fref{criterioncmfpt}. The plot gives a phase diagram in terms of the parameters from which we can identify the regions in which $CV_+ - \Lambda_+>0$  which indicates that resetting can minimize
$\langle t_\lambda\rangle_+$. On the other hand, we also identify regions where $CV_+ - \Lambda_+<0$, and thus there exists no optimal resetting rate which can minimize $\langle t_\lambda\rangle_+$ in this parameter regime. A similar analysis can also be done for the conditional exit through the other boundary. In effect, the criterion (\ref{Condition for optimality-conditional}) allows us to identify wedge configurations that can help decide whether to utilize resetting for a selective or non-selective outcome.

\section{conclusion}\label{section6}
In this paper, we focused on understanding how diffusion, when influenced by stochastic resetting, behaves within a geometrically non-trivial environment—specifically, the wedge domain. While diffusion in a wedge with absorbing boundaries is a well-established problem, with first-passage statistics known to be strongly affected by the system’s geometry, our central aim was to explore how the introduction of a Poissonian resetting mechanism modifies these statistics. We first analyzed the unconditional first-passage behavior and demonstrated that the mean first-passage time to the boundaries can be minimized by appropriately tuning the resetting rate. This result was further validated through extensive numerical simulations, confirming the optimization effect introduced by resetting dynamics.

To assess whether stochastic resetting optimizes mean first-passage times, we evaluated the coefficient of variation (CV) for different system parameters. Our analysis reveals that the radial parameters effectively scale out of the problem. The CV analysis also shows that not all geometric configurations of the wedge domain lead to optimal behavior, a result that has been corroborated by numerical simulations. This suggests a behavioral transition, wherein the mean exit times may or may not be optimized depending on the specific configuration of the wedge domain. The phase diagram highlights interesting regimes where, even for small wedge angles, there exist resetting angles at which diffusion becomes a less dominant mechanism in the particle absorption process at the boundaries. The exact formula obtained for the coefficient of variation (CV) allows one to identify several quantitative features for the phase diagram, such as the critical angle $\alpha^{max}_c=\cos^{-1}(\frac{1+\sqrt{55}}{9})$ above which resetting always results in optimal mean first-passage time. It will be interesting to explore the existence of Landau-like first-order crossovers in this system, similar to those observed in other models \cite{Pal-VVP-PRR2019}.

To investigate the impact of stochastic resetting on first-passage behavior under conditioned absorption, we examine the flux of the system through its boundaries for wedge angles that are integer divisions of $\pi$. By utilizing expressions for current densities in the Laplace space, we analytically trace the behavior of conditioned mean first-passage times and splitting probabilities. Our analysis reveals that the conditioned mean first-passage times attain a minimum at a non-zero reset rate, indicating that the process benefits from resetting dynamics. The speed-up in escape rate through one of the boundaries is shown to depend on both the geometry of the system and the resetting rate. Furthermore, the preferential exit is enhanced by the reset rate, particularly when the process is initiated asymmetrically. This suggests that the preferential absorption is influenced not only by the system's geometric properties but also by the external resetting dynamics. In order to determine if a non-zero reset benefits conditioned exit timings, we also conducted an analysis on the optimality criterion for conditional exit times, which is a sufficient but not necessary condition.

As a potential direction for future research, it would be interesting to extend the dynamics of diffusion within a wedge geometry by incorporating an additional convective term \cite{krapviskyApexflow} to investigate the interplay between convection and stochastic resetting, following a similar approach to those presented in \cite{ray2019peclet,biswas2023rate}. Additionally, this study could be extended to a fully confined wedge geometry 
$\{(r,\theta)\colon 0<\theta<\alpha,\; 0<r\le L\},$ with absorbing boundary conditions (Dirichlet) on the straight edges \(\theta=0\) and \(\theta=\alpha\), and a reflecting (Neumann) circular boundary at \(r=L\) \cite{baravi2025solutions}.
 Another promising avenue would be to study the first-passage properties under threshold (described by one of the edges) resetting introduced in \cite{biswas2025target}.

\medskip
\section{ACKNOWLEDGMENTS}
FN and VVP acknowledge SERB Start-up Research Grant No.~2022/001077 for research support. AP acknowledges research support from the Department of Atomic Energy via the SoftMatter Apex projects. AP also acknowledges the International Research Project (IRP) titled ``Classical and quantum dynamics in out of equilibrium systems'' by CNRS, France.
\bibliography{bibrefV9.bib}
%\addbibresource{bibref.bib}
%\bibliography{apssamp}
%\bibliographystyle{aipauth4-1}
%\printbibliography
\newpage

\begin{titlepage}
\centering
{\Large \underline{Appendix} \\}
\vspace{0.3 cm}
{\Large ``Diffusion in a wedge geometry: First-Passage Statistics under Stochastic Resetting''\\}
\vspace{0.5 cm}
\end{titlepage}

\onecolumngrid
\setcounter{page}{1}
\renewcommand{\thepage}{A\arabic{page}}
\setcounter{equation}{0}
\renewcommand{\theequation}{A\arabic{equation}}
\setcounter{figure}{0}
\renewcommand{\thefigure}{A\arabic{figure}}
\setcounter{section}{0}
\renewcommand{\thesection}{A\arabic{section}}
\setcounter{table}{0}
\renewcommand{\thetable}{A\arabic{table}}
\setcounter{subsection}{0}
\renewcommand{\thesubsection}{\alph{subsection}}

\noindent 
%This Supplemental Material 
Here we provide supporting mathematical derivations of main results described in the article and elaborate on the numerical simulation methods used to generate the data to verify the analytics.

%\tableofcontents

% \appendix
% \begin{widetext}
 
\section{Derivation of probability density function for $\alpha=\pi/n$ \eref{dypdfexpression1}}\label{pdf_pibyn}
We present here a derivation of the expression of the PDF (\eref{dypdfexpression1}), which was earlier obtained in \cite{dy}. Here we provide the expression as a special case: $\alpha=\pi/n$, ($n$, being positive integer) of the general time-domain solution \eref{pdfgenerictimedomain} for the probability density of a diffusing particle inside a two-dimensional wedge of apex angle $\alpha$. The general expression of the PDF is
\begin{equation}\label{A1} P_0(r,\theta,t) = \frac{\exp\big(-\frac{r^2+r_0^2}{4Dt}\big)}{\alpha D t} \sum_{m=1}^{\infty}\sin\Big(\frac{m\pi\theta_0}{\alpha}\Big)\sin\Big(\frac{m\pi\theta}{\alpha}\Big) I_{\frac{m\pi}{\alpha}}\Big(\frac{rr_0}{2Dt}\Big), \end{equation} 
where $I_\nu$ denotes the modified Bessel function of the first kind. A particle undergoing Brownian motion initiates the dynamics at position $({r_0,\theta_0}) $ and $P(r,\theta,t)$ gives the measure of probability density at an arbitrary location $(r,\theta)$  at time $t$ inside the wedge domain($0\le\theta\le\alpha,\;0\le r\le\infty$).

For the derivation of $P(r,\theta,t)$  for the special case of $\alpha=\frac{\pi}{n},\quad n\in\mathbb{Z}{>0}$  [\eref{dypdfexpression1}]. We first expand the sine terms inside the series using  the elementary identity $\sin A\sin B=\tfrac{1}{2}[\cos(A-B)-\cos(A+B)]$ and rewrite the series as
\begin{equation}\label{trigsersup}
    \sum_{m=1}^{\infty}\sin\Big(\frac{m\pi\theta_0}{\alpha}\Big)\sin\Big(\frac{m\pi\theta}{\alpha}\Big) I_{\frac{m\pi}{\alpha}}\Big(\frac{rr_0}{2Dt}\Big)=\frac{1}{2}\sum_{m=1}^{\infty} \Big(\cos(n m(\theta_0-\theta))-\cos(n m (\theta_0+\theta))\Big)I_{n m}\Big(\frac{rr_0}{2Dt}\Big)
\end{equation}
 where we have replaced $\alpha=\pi/n$ and with the shorthand notation
\begin{equation}\label{defs}
x = \theta_0-\theta,\qquad y = \theta_0+\theta,\qquad z = \frac{rr_0}{2Dt}, 
\end{equation} 
The RHS of \eqref{trigsersup} can be compactly written as
\begin{equation}\label{Sndef}
\begin{split}
\frac{1}{2}\sum_{m=1}^{\infty} \Big(\cos(n m x)-\cos(n m y)\Big)I_{n m}(z) 
=\tfrac{1}{2}\big(S_n(x,z)-S_n(y,z)\big), 
\end{split}
\end{equation} where we identify the full series as sum of two identical series $S_n(x,z)\equiv\sum_{m=1}^{\infty} \cos(n m x)I_{n m}(z)\;\text{and}\;S_n(y,z)\equiv\sum_{m=1}^{\infty} \cos(n m y)I_{n m}(z)$. One can clearly note that the two split-sums are mathematically equivalent, therefore we can define a common representative term as
\begin{equation}\label{Sn_def} S_n(u,z)=\sum_{m=1}^{\infty}\cos(n m u)I_{n m}(z),\qquad u\in{x,y}. \end{equation}

\noindent and complete the derivation with the evaluation of $S_n(u,z)$.
Let's begin by putting $j=mn$. Then $j$ ranges over the positive integers that are multiples of $n$, i.e. $j\in{n,2n,3n,\dots}$, therefore \begin{equation}\label{Sn_reindex} 
S_n(u,z)=\sum_{\substack{j\ge 1\ \\ j\equiv 0(\mathrm{mod}\;n)}} I_j(z)\cos(ju). \end{equation}

As indicated in the summation above \eqref{Sn_reindex}, the series expands to only those $j's$ that are integer multiples of $n$. To generalize the series and isolate the subsequence $j\equiv 0(\mathrm{mod}\;n)$ we use the sum of $n^{th}$ roots of unity. Let $\omega=e^{2\pi i/n}$. For any integer $j$ 
\begin{equation}\label{roots_unity} 
\frac{1}{n}\sum_{k=0}^{n-1}\omega^{kj} = \begin{cases} 1, & j\equiv 0(\mathrm{mod}\;n), \\ 0 &\text{otherwise.} \end{cases} 
\end{equation} 
Multiplying by $e^{i j u}$ and taking the real part gives the corresponding projection for cosines:

\begin{equation}\label{proj-cos} \frac{1}{n}\sum_{k=0}^{n-1}\cos\Big(j\Big(u+\frac{2\pi k}{n}\Big)\Big) =\begin{cases}\cos(j u), & j\equiv 0(\mathrm{mod}\;n),\\0 & \text{otherwise.}\end{cases} 
\end{equation} 

Applying \eqref{proj-cos} term wise to the full (unrestricted) series $\sum_{j=1}^\infty I_j(z)\cos(ju)$ filters out exactly those $j$ divisible by $n$. Hence 

\begin{equation}\label{Sn_projection}
S_n(u,z)=\frac{1}{n}\sum_{k=0}^{n-1}\sum_{j=1}^\infty I_j(z)\cos\Big(j\Big(u+\frac{2\pi k}{n}\Big)\Big).
\end{equation}

The inner infinite series in \eqref{Sn_projection}, can be closed using the familiar generating identity for modified Bessel function of the first kind $I_j(z)$(valid for complex $z$ and real $t$) \begin{equation}\label{gen-I} e^{z\cos t}=I_0(z)+2\sum_{j=1}^\infty I_j(z)\cos(jt). \end{equation} From \eqref{gen-I} we have 
\begin{equation}\label{sum_Icos} \sum_{j=1}^\infty I_j(z)\cos(jt)=\frac{e^{z\cos t}-I_0(z)}{2}.
\end{equation} Set $t=u+\tfrac{2\pi k}{n}$ in \eqref{sum_Icos} and substitute into \eqref{Sn_projection}. Summing over $k$ produces \begin{equation}\label{Sn_intermediate} \begin{aligned} S_n(u,z) &= \frac{1}{n}\sum_{k=0}^{n-1}\frac{e^{z\cos\big(u+\tfrac{2\pi k}{n}\big)}-I_0(z)}{2} = \frac{1}{2n}\sum_{k=0}^{n-1} e^{z\cos\big(u+\tfrac{2\pi k}{n}\big)} -  \frac{1}{2}I_0(z). \end{aligned} \end{equation}

Note that in the difference $S_n(x,z)-S_n(y,z)$ the constant contributions $-\tfrac{1}{2}I_0(z)$ cancel, so only the exponential sums generated by \eqref{gen-I} remain.

Substitute \eqref{Sn_intermediate} (for $u=x$ and $u=y$) into \eqref{Sndef}, and then insert the result into \eqref{A1}. Using the definitions \eqref{defs} and simplifying the exponential factors yields 
\begin{equation}\label{final_pdf} 
\begin{aligned} P_0(r,\theta,t) &=\frac{\exp\big(-\tfrac{r^2+r_0^2}{4Dt}\big)}{\tfrac{\pi}{n} D t}\cdot\frac{1}{4n} \sum_{k=0}^{n-1}\Bigg(e^{z\cos\big(x+\tfrac{2\pi k}{n}\big)}-e^{z\cos\big(y+\tfrac{2\pi k}{n}\big)}\Bigg)\\&=\frac{1}{4\pi Dt}\sum_{k=0}^{n-1}\Bigg( \exp\Big(-\frac{r^2+r_0^{2}-2 r r_0\cos\big(\tfrac{2\pi k}{n}+\theta_0-\theta\big)}{4Dt}\Big)\ -\exp\Big(-\frac{r^2+r_0^{2}-2 r r_0\cos\big(\tfrac{2\pi k}{n}+\theta_0+\theta\big)}{4Dt}\Big)\Bigg),
\end{aligned} 
\end{equation} 
which is equivalent to \eref{dypdfexpression1} used in the main text. 
The form of finite sum involving the second exponential in \eref{dypdfexpression1} can be obtained by rearranging the second exponential sum in \eref{final_pdf} as following
\begin{equation}\label{comparisontody1}
\sum_{k=0}^{n-1}e^{z\cos\Big(\tfrac{2\pi k}{n}+y\Big)}=\sum_{k=0}^{n-1}e^{z\cos\Big(\tfrac{2(n-k)\pi}{n}-y\Big)}
  \end{equation}
where we have used the fact $\cos(\theta)=\cos(2\pi-\theta)$. Now by reindexing $n-k=j$, which reverses the order of the terms, and relabeling $j=k+1$  results in
\begin{equation}\label{comparisontody2}
\sum_{k=0}^{n-1}e^{z\cos\big(2(n-k)\pi /n-y\big)}=\sum_{j=1}^{n} e^{\,z\cos(2j\pi/n-y)}=\sum_{k=0}^{n-1} e^{\,z\cos(2(k+1)\pi/n-y)}
  \end{equation}
which is identical to the second sum in \eref{dypdfexpression1}.

\section{Derivation of Laplace transform of the survival probability \eref{survivallaplacedefsum}}\label{survival_laplace}
We derive the Laplace transform (time variable \(t\mapsto s\)) of the survival probability given in \eref{survivallaplacedefsum}.  We start from the alternate time–space representation (see \cite{majumder} for a complete derivation of the time-space representation)
\begin{equation}\label{survivalmodifiedform}
\begin{split}
Q_0(r_0,\theta_0,t)
={}&\; \operatorname{Erf}\!\Big( \sqrt{2z_0}\,\sin\!\big(\min\{\theta_0,\tfrac{\pi}{2}\}\big) \Big) \\
&\; + \sum_{j=1}^{\frac{\pi}{2\alpha} + \frac{1}{2}} (-1)^j
  \Bigg[
    \operatorname{Erf}\!\Big(\sqrt{2z_0}\,\sin\!\big(\min\{j\alpha+\theta_0,\tfrac{\pi}{2}\}\big)\Big)
    - \operatorname{Erf}\!\Big(\sqrt{2z_0}\,\sin\!\big(\min\{j\alpha-\theta_0,\tfrac{\pi}{2}\}\big)\Big)
  \Bigg] \\
&\; + \tfrac{1}{2}\Big(\tfrac{2}{\pi}\Big)^{3/2} \sqrt{z_0}\, e^{-z_0}
     \int_{0}^{\infty} e^{-z_0\cosh u}\,\sinh\!\Big(\tfrac{u}{2}\Big)
     \Bigg[
      \arctan\!\Big(\frac{\sin\!\big(\tfrac{\pi}{\alpha}(\theta_0 + \tfrac{\pi}{2})\big)}
                      {\sinh\!\big(\tfrac{\pi u}{2\alpha}\big)}\Big)
      + \arctan\!\Big(\frac{\sin\!\big(\tfrac{\pi}{\alpha}(\theta_0 - \tfrac{\pi}{2})\big)}
                      {\sinh\!\big(\tfrac{\pi u}{2\alpha}\big)}\Big)
   \Bigg]\,du,
\end{split}
\end{equation}
where we have used the standard short-hand
\[
z_0 \equiv \frac{r_0^2}{8Dt}.
\]

Because the Laplace transform acts only on the temporal variable \(t\), it is convenient to split \(Q_0\) into two parts: the finite sum of error-function terms (which we shall denote by \(G(t)\)) and the remaining integral term (denoted by \(H(t)\)).  Specifically
\begin{equation}\label{GHdef}
\begin{aligned}
Q_0(r_0,\theta_0,t) &= G(t) + H(t),\\[4pt]
G(t) &= \operatorname{Erf}\!\Big(\sqrt{2z_0}\,\sigma_0\Big)
+ \sum_{j=1}^{\frac{\pi}{2\alpha}+\frac12} (-1)^j\Big[
\operatorname{Erf}\!\big(\sqrt{2z_0}\,\sigma_{+,j}\big)
- \operatorname{Erf}\!\big(\sqrt{2z_0}\,\sigma_{-,j}\big)
\Big],\\[4pt]
H(t) &= \int_{0}^{\infty} F(\theta_0,u)\,\sqrt{z_0}\,\exp\!\big(-z_0(1+\cosh u)\big)\,du,
\end{aligned}
\end{equation}
with the following symbols \(\sigma\in\{\sigma_0,\sigma_{\pm,j}\}\),
\[\sigma_0=\sin\!\big(\min\{\theta_0,\tfrac{\pi}{2}\}\big),\qquad
\sigma_{+,j}=\sin\!\big(\min\{j\alpha+\theta_0,\tfrac{\pi}{2}\}\big),\qquad
\sigma_{-,j}=\sin\!\big(\min\{j\alpha-\theta_0,\tfrac{\pi}{2}\}\big),
\]
and where the function \(F(\theta_0,u)\) collects the remaining \(u\)-dependent prefactors that appear in the third term of \eqref{survivalmodifiedform} (i.e. all factors that do not multiply the \(\sqrt{z_0}e^{-z_0(1+\cosh u)}\) kernel).
\vspace{6pt}
\subsection{Laplace transform of the error–function terms \(G(t)\)}

We compute the Laplace transform, \(\mathcal{L}\{G\}(s)\) term-by-term.  The Laplace transform acts on each prototype form of Erf function, written in the form \(\operatorname{Erf}\big(A/\sqrt{t}\big)\), where \(A\) is defined
, with \(z_0=r_0^2/(8Dt)\) as,
\begin{equation}
    \sqrt{2z_0}\,s = \sqrt{\frac{2r_0^2}{8Dt}}\sigma =
\frac{1}{\sqrt{t}}\underbrace{\Big(\frac{r_0}{2\sqrt{D}}\sigma\Big)}_{A},
\end{equation}

with an appropriate constant \(A\) (for example, for the first term \(A=A_0=\tfrac{r_0}{2\sqrt{D}}\sin(\min\{\theta_0,\tfrac{\pi}{2}\})\) and or \(A=A_{\pm,j}\in\{\tfrac{r_0}{2\sqrt{D}}\sin\!\big(\min\{j\alpha\pm\theta_0,\tfrac{\pi}{2}\}\big)\)\}).
\vspace{10pt}
\\
We define the Laplace transform $\mathcal{L}\{F(t)\}(s)$
\begin{equation}
\begin{aligned}
\mathcal{L}\{\operatorname{Erf}(A/\sqrt{t})\}(s)
&= \int_0^\infty e^{-st}\, \operatorname{Erf}\!\Big(\frac{A}{\sqrt{t}}\Big)\,dt \\
&= \frac{2}{\sqrt{\pi}}\int_0^\infty e^{-st}\Big(\int_0^{A/\sqrt{t}} e^{-y^2}\,dy\Big)\,dt,
\end{aligned}
\end{equation}
using the representation \(\operatorname{Erf}(u)=\tfrac{2}{\sqrt{\pi}}\int_0^u e^{-y^2}dy\).  Since the integrand is non-negative we may apply Fubini's theorem and swap the order of integration; the integration domain \(\{(t,y):t>0,0\le y\le A/\sqrt{t}\}\) is equivalent to \(\{(y,t):y\ge0,0\le t\le A^2/y^2\}\).  Carrying out the \(t\)-integral first gives
\begin{equation}\label{interchangeintegrals}
\begin{aligned}
\mathcal{L}\{\operatorname{Erf}(A/\sqrt{t})\}(s)
&= \frac{2}{\sqrt{\pi}}\int_0^\infty e^{-y^2}\Big(\int_0^{A^2/y^2} e^{-st}\,dt\Big)\,dy \\
&= \frac{2}{s\sqrt{\pi}}\int_0^\infty e^{-y^2}\Big(1-e^{-sA^2/y^2}\Big)\,dy \\
&= \frac{2}{s\sqrt{\pi}}\Big[\int_0^\infty e^{-y^2}\,dy - \int_0^\infty e^{-y^2-sA^2/y^2}\,dy\Big].
\end{aligned}
\end{equation}
\vspace{10pt}
\\
The first integral is a non-normalized Gaussian:
\[
\int_0^\infty e^{-y^2}\,dy = \frac{\sqrt{\pi}}{2}.
\]
The second integral can be evaluated using the formula
\[
\int_0^\infty e^{-\alpha y^2 - \beta/y^2}\,dy
= \tfrac{1}{2}\sqrt{\frac{\pi}{\alpha}}\,e^{-2\sqrt{\alpha\beta}}
\qquad(\textbf{Re}(\alpha),\textbf{Re}(\beta)>0)
\]
With \(\alpha=1\) and \(\beta=sA^2\) this yields,
\begin{equation}
\int_0^\infty e^{-y^2 - \frac{sA^2}{y^2}}\,dy = \frac{\sqrt{\pi}}{2}\,e^{-2A\sqrt{s}}.
\end{equation}
Combining these results gives the closed form
\begin{equation}
 \mathcal{L}\{\operatorname{Erf}(A/\sqrt{t})\}(s) = \frac{1}{s}\big(1 - e^{-2A\sqrt{s}}\big).
\end{equation}
\vspace{10pt}
\\
We can now apply the transform to all the Erf terms with \(A\) replaced by the relevant constants, by linearity of the Laplace transform, we obtain the contribution of the entire \(G(t)\) block:
\begin{equation}\label{erfpart}
\begin{split}
\mathcal{L}\{G\}(s)
&= \frac{1}{s}\Big(1 - e^{-2A_0\sqrt{s}}\Big)
+ \sum_{j=1}^{\frac{\pi}{2\alpha}+\frac12} \frac{(-1)^j}{s}\Big(e^{-2A_{-,j}\sqrt{s}} - e^{-2A_{+,j}\sqrt{s}}\Big)\\[4pt]
&= \frac{1}{s}\Bigg[1 - e^{-2A_0\sqrt{s}}
+ \sum_{j=1}^{\frac{\pi}{2\alpha}+\frac12} (-1)^j\big(e^{-2A_{-,j}\sqrt{s}} - e^{-2A_{+,j}\sqrt{s}}\big)\Bigg],
\end{split}
\end{equation}
with \(A_0\) and \(A_{\pm,j}\) defined as above.

\vspace{6pt}
\subsection{Laplace transform of the integral term \(H(t)\)}

\noindent Let's switch to following notations
\begin{equation}\label{defr}
\tilde{r}=\frac{r_0^2}{8D},\qquad z_0=\frac{\tilde{r}}{t}.
\end{equation}
With this substitution the integral term in \eqref{GHdef} may be written as
\begin{equation}
H(t)=\sqrt{\tilde{r}}\,t^{-1/2}\int_{0}^{\infty} F(\theta_0,u)\,e^{-\frac{\tilde{r}(1+\cosh u)}{t}}\,du.
\end{equation}
The Laplace transform of \(H(t)\) is obtained by interchanging the order of the \(t\)- and \(u\)-integrals (justified by Fubini's theorem, since the integrand decays sufficiently fast for Re\((s)>0\)) and then evaluating the resulting \(t\)-integral(Due to the Laplace transform) using the identity
\begin{equation}\label{basset-identity}
\int_0^\infty t^{\nu-1} e^{-\beta t - \gamma/t}\,dt
=2\Big(\frac{\gamma}{\beta}\Big)^{\nu/2}K_{\nu}\!\big(2\sqrt{\beta\gamma}\big),
\end{equation}
valid for Re\((\beta)\),Re\((\gamma)>0\). For our choice
\[
\nu=\frac12,\qquad \beta=s,\qquad \gamma=\tilde{r}(1+\cosh u).
\]
and using \(K_{1/2}(w)=\sqrt{\pi/(2w)}\,e^{-w}\) we obtain the intermediate identity
\begin{equation}
    \int_0^\infty t^{-1/2} e^{-st - \frac{\tilde{r}(1+\cosh u)}{t}}\,dt
= \sqrt{\frac{\pi}{s}}\,\exp\!\big(-2\sqrt{s\tilde{r}(1+\cosh u)}\big).
\end{equation}
Noting that \(1+\cosh u = 2\cosh^2(u/2)\) and substituting \(\tilde{r}=r_0^2/(8D)\) yields after minor simplification the following closed form for \(\mathcal{L}\{H\}(s)\):
\begin{equation}\label{intpart}
    \mathcal{L}\{H\}(s)
= \frac{r_0\sqrt{\pi}}{2\sqrt{2D}}\;\frac{1}{\sqrt{s}}\;
\int_{0}^{\infty} F(\theta_0,u)\;
\exp\!\Big(-r_0\sqrt{\frac{s}{D}}\cosh\!\frac{u}{2}\Big)\,du.
\end{equation}

\bigskip
Finally, combining \eqref{erfpart} and \eqref{intpart} yields the Laplace-domain representation \(\mathcal{L}\{Q_0\}(s)=\mathcal{L}\{G\}(s)+\mathcal{L}\{H\}(s)\), which is algebraically identical to the expression given in \eref{survivallaplacedefsum}. This completes the derivation.

\section{Alternate expression for Laplace transform of survival probability \eref{survivaltimespace} and derivation of first moment of first-passage time (\eref{firstmomentzeroreset})}\label{laplace_surv}
We evaluate the Laplace transform of the series expression for the survival probability \eref{survivaltimespace} using $Mathematica$ software and arrive at the following expression for the Laplace transform
\begin{equation}\label{altsurv}
    \begin{split}
        q_0(r_0,\theta_0,s)=\frac{2\pi}{\alpha} \sum _{m=0}^{\infty } \csc \left(u_m \pi\right) \sin      \left(2u_m \theta_0\right) \left[\frac{\tilde{r}}{u_m} \underbrace{\left(_1\tilde{F}_2\left(1;1-u_m,2+u_m;\tilde{r}s\right)- _1\tilde{F}_2\left(1;2-u_m,1+u_m;\tilde{r}s\right)\right)}_{A_m(s)}+\underbrace{\frac{I_{2u_m}\left(2\sqrt{2\tilde{r}s}\right)}{s}}_{B_{m}(s)}\right].
    \end{split}
\end{equation}
where $q_0(r_0,\theta_0,s)=\mathcal{L}\{Q_0(r_0,\theta_0,t\}(s)$, $\tilde{r}=z_0t$ (defined in \eqref{defr}) and $u_m= \frac{(2m+1)\pi}{2\alpha}.$ The notation $_1\tilde{F}_2(a;b,c;d)$ stands for the regularized Hypergeometric function. The term $\csc(u_m\pi)$ in the given form  of $q_0(r_0,\theta_0,s)$, appears to make the whole expression singular for $\alpha=\pi/n\;\text{for}\;n\;\text{is even}$. However, a close analysis shows that this is a removable singularity and evaluation for $n=$ even cases require the appropriate use of L'Hopital's rule. Due to which the singularity is explicitly removed when $\lim_{s\rightarrow 0} q_0(r_0,\theta_0,s)$ is evaluated, and is demonstrated below (\textbf{{\ref{limitofhyper}}}).
\subsection{Expression of First moment of the first-passage time without reset}
We use this expression (\eref{altsurv}) of the survival probability to derive the expression for first moment of the first-passage time (\eref{mfptseries}). In Laplace transformed space, the mean first-passage time $\langle t_0\rangle$, is defined as
\begin{align}\label{mfptseries}
    \nonumber\langle t_0\rangle&=\lim_{s\rightarrow 0}q_0(r_0,\theta_0,s)\\
    &=\frac{2\pi}{\alpha} \sum _{m=0}^{\infty } \csc \left(u_m \pi\right) \sin\left(2u_m \theta_0\right)\left(\frac{\tilde{r}}{u_m}\lim_{s\rightarrow 0}A_m(s)+\lim_{s\rightarrow 0}B_m(s)\right),
\end{align}
where the limits will be analyzed in the following. 
%DEFINE A_m and B_m separately here.
\subsubsection{$\lim_{s\rightarrow 0}A_m(s)$}\label{limitofhyper}
To analyze this limit, we first express the regularized Hypergeometric function $_1\tilde{F}_2\left(1;b,c;z\right)$ in terms of the generalized Hypergeometric function
\begin{equation}\label{hypergeoexpansion}
    _1\tilde{F}_2\left(1;b,c;z\right)=\frac{_1F_2\left(1;b,c;z\right)}{\Gamma(b)\Gamma(c)}
\end{equation}
Here $\Gamma(w)$ is the Euler gamma function and since for any generalized Hypergeometric function $_1F_2\left(1;b,c;z\right)\sim 1+\mathcal{O}(z)$, application of limit on \eqref{hypergeoexpansion} yields
\begin{equation}
\begin{split}\label{limitingbehavior}
    \lim_{z\rightarrow 0}\;_1\tilde{F}_2\left(1;b,c;z\right) &=\frac{1}{\Gamma(b)\Gamma(c)}
\end{split}
\end{equation}
Therefore 
\begin{equation}
\begin{split}
    \lim_{s\rightarrow 0} A_m(s)&=\lim_{s\rightarrow0}\left(_1\tilde{F}_2\left(1;1-u_m,2+u_m;\tilde{r}s\right)- _1\tilde{F}_2\left(1;2-u_m,1+u_m;\tilde{r}s\right)\right)\\&=\frac{1}{\Gamma(1-u_m)\Gamma(2+u_m)}-\frac{1}{\Gamma(2-u_m)\Gamma(1+u_m)}
\\ &=\frac{1}{\Gamma(1-u_m)(1+u_m)u_m\Gamma(u_m)}-\frac{1}{(1-u_m)(-u_m)\Gamma(-u_m)\Gamma(1+u_m)}
\\ &=\frac{2}{\pi}\frac{\sin(\pi u_m)}{(u_m^2-1)} \qquad{(\text{Using reflection identity}\;\Gamma(w)\Gamma(1-w)=\pi/\sin(\pi w))}
\end{split}
\end{equation}
The term $\sin(\pi u_m)$ exactly cancels $\csc(\pi u_m)$ in \eref{mfptseries}.
\subsubsection{$\lim_{s\rightarrow 0}B_m(s)$}
Using the asymptotic behavior($z\rightarrow 0$) of Modified Bessel of the first kind defined in \eqref{besselasymptotic}
\begin{align}\label{bessellimits}
    \nonumber\lim_{s\rightarrow 0}B_m(s)&= \lim_{s\rightarrow 0}\frac{I_{2u_m}\left(2\sqrt{2\tilde{r}s}\right)}{s}\\
    &= 8{\tilde{r}}\lim_{z\rightarrow 0} \frac{I_\mu(\sqrt{z})}{z}
    \sim\lim_{z\rightarrow 0}\frac{2^{-\mu}}{\Gamma(\mu+1)}z^{\frac{\mu}{2}-1}
\end{align}
we see the behavior of this limit is governed by the exponent $\frac{\mu}{2}-1$. The following distinct cases arise:(1) $\mu/2-1>0$ and (2) $\mu/2-1\leq0$. We look into these cases by re-substituting  $\mu=2u_m$
\begin{itemize}
    \item \textbf{\textit{Case 1:} $\alpha< (2m+1)\pi/2$}: The limit \eqref{bessellimits} vanishes for all $m\geq0$. 
    \item \textbf{\textit{Case 2:} $\alpha>(2m+1)\pi/2$}: This limit diverges for all $m$, hence the mean first-passage time diverges for all wedge angles \hspace{5em}$\alpha>\pi/2 (m=0)$ .
    \item \textbf{\textit{Case 3:} $\alpha=(2m+1)\pi/2$}: This makes the asymptotic dependence of $I_\mu(z)$ independent of z with the limit converging to a constant value ($r_0^2/8D$) which is independent of $m$. Also this fixes $u_m=1$ which implies that the second part of the series \eqref{mfptseries} ($\sum_{m=0}^{\infty}\csc(\pi)\sin(2\theta_0)\lim_{s\rightarrow 0}B_m(s)$) diverges. Thus $\langle t_0\rangle\rightarrow\infty$, for all $\alpha\geq\pi/2.$
\end{itemize}

\par
\noindent The above cases fixes the finite mean first-passage domain,\textit{i.e,} $\alpha\in[0,\pi/2)$. In this domain, we can plug in the corresponding limiting behaviors of $A_m(s)$ and $B_m(s)$ into \eqref{mfptseries} and obtain the expression for $\langle t_0\rangle$
\begin{equation}\label{derivedmfpt}
     \langle t_0\rangle=\frac{4\tilde{r}}{\alpha} \sum _{m=0}^{\infty }\left(\frac{\sin\left(2u_m \theta_0\right)}{u_m(u_m^2-1)}\right)
\end{equation}
Re-substituting back for $u_m$ and $\tilde{r}$ in \eref{derivedmfpt}, we obtain the original expression for $\langle t_0 \rangle$\eqref{mfptseries}. 
\par The expression for the second moment $\langle t_0^2\rangle$\eqref{secondmomentzeroreset},which is defined as
\begin{equation}
    \langle t_0^2\rangle=\lim_{s\rightarrow 0}\left(-2\frac{\partial q_0}{\partial s}\right)
\end{equation}
can also be derived following a similar procedure of evaluating the limits.
\section{Derivation of closed form expressions for first moment ($\langle t_0 \rangle$), second moment ($\langle t_0^2\rangle$) and the coefficient of variation $CV$.}\label{closedforms}
\subsection{\textbf{Derivation of the closed form of the First moment}}\label{firmomclosed}
\noindent We present here, a derivation for the closed form of the series expression for the first moment $\langle t_0 \rangle$
\begin{equation}\label{series-proto}
\langle t_0 \rangle=\frac{4r_0^2\alpha^2}{\pi D}\sum_{m=0}^{\infty}
\frac{\sin\!\big(\tfrac{(2m+1)\pi\theta_0}{\alpha}\big)}
{(2m+1)\big((2m+1)^2\pi^2-4\alpha^2\big)}.
\end{equation}

For the derivation below we assume
\[
0<\theta_0<\alpha,\qquad \alpha\not\in\Big\{\frac{(2n+1)\pi}{2}:n\in\mathbb Z\Big\},
\]
so that the denominators in \eqref{series-proto} are nonzero.

\noindent Let's introduce the odd index \(k=2m+1\). Also for brevity, we define the dimensionless variables
\begin{equation}\label{xddef}
x=\frac{\pi\theta_0}{\alpha},\qquad \Lambda=\frac{2\alpha}{\pi}.
\end{equation}
Note that with \(0<\theta_0<\alpha\) we have \(0<x<\pi\) and \(\Lambda>0\). 
\par 
\noindent Rewriting \eqref{series-proto} in terms of \(k\) yields
\begin{equation}
\langle t_0 \rangle=\frac{4r_0^2\alpha^2}{\pi D}\sum_{\substack{k}}
\frac{\sin(kx)}{k\big(k^2\pi^2-4\alpha^2\big)}
= \frac{4r_0^2\alpha^2}{\pi D}\cdot\frac{1}{\pi^2}\sum_{\substack{k}}
\frac{\sin(kx)}{k\,(k^2-\Lambda^2)}.
\end{equation}
Thus it suffices to evaluate the series
\begin{equation}\label{Sdef}
S(x,\Lambda)=\sum_{\substack{k}}\frac{\sin(kx)}{k\,(k^2-\Lambda^2)},
\qquad 0<x<\pi,\ \Lambda\notin\mathbb Z_{\text{odd}}.
\end{equation}

\ Differentiate the series term wise twice:
\[
S'(x)=\sum_{k}\frac{\cos(kx)}{k^2-\Lambda^2},\qquad
S''(x)=-\sum_{k}\frac{k\sin(kx)}{k^2-\Lambda^2}.
\]
Form the combination \(S''+\Lambda^2 S\):
\begin{equation}
\begin{aligned}
S''(x)+\Lambda^2 S(x)
&= -\sum_{k}\frac{k\sin(kx)}{k^2-\Lambda^2}
+ \Lambda^2\sum_{k}\frac{\sin(kx)}{k(k^2-\Lambda^2)}\\[4pt]
&= -\sum_{k}\frac{(k^2-\Lambda^2)\sin(kx)}{k(k^2-\Lambda^2)}
= -\sum_{k}\frac{\sin(kx)}{k}.
\end{aligned}
\end{equation}
For \(0<x<\pi\) the Fourier identity of odd sines($k=1,3,5...$) for the square wave gives,
\begin{equation}
    \sum_{\substack{k}}\frac{\sin(kx)}{k}=\frac{\pi}{4}.
\end{equation}
Hence \(S\) satisfies the constant-coefficient ODE
\begin{equation}\label{Sode}
S''(x)+\Lambda^2 S(x) = -\frac{\pi}{4},\qquad 0<x<\pi.
\end{equation}

The general solution of \eqref{Sode} is
\[
S(x)=C\cos(\Lambda x)+D\sin(\Lambda x)-\frac{\pi}{4\Lambda^2}.
\]
The boundary behavior of the original series yields \(S(0)=0\) and \(S(\pi)=0\) (these follow from the factor \(\sin(kx)\) vanishing at \(x=0,\pi\)). Imposing \(S(0)=0\) gives
\[
C-\frac{\pi}{4\Lambda^2}=0 \quad\Rightarrow\quad C=\frac{\pi}{4\Lambda^2}.
\]
and \(S(\pi)=0\) gives
\[
\frac{\pi}{4\Lambda^2}\cos(\Lambda\pi)+D\sin(\Lambda\pi)-\frac{\pi}{4\Lambda^2}=0,
\]
so
\[
D=\frac{\pi}{4\Lambda^2}\,\frac{1-\cos(\Lambda\pi)}{\sin(\Lambda\pi)}
=\frac{\pi}{4\Lambda^2}\tan\!\Big(\frac{\Lambda\pi}{2}\Big),
\]
where the last equality uses the trigonometric identity
\(\dfrac{1-\cos y}{\sin y}=\tan(y/2)\) with \(y=\Lambda\pi\).

Therefore the unique solution on \(0<x<\pi\) is
\begin{equation}\label{Sclosed}
S(x,\Lambda)=\frac{\pi}{4\Lambda^2}\Big(-1+\cos(\Lambda x)+\tan\!\Big(\frac{\Lambda\pi}{2}\Big)\sin(\Lambda x)\Big).
\end{equation}

Use \eqref{xddef}: \(\Lambda x = 2\theta_0\) and \(\tfrac{\Lambda\pi}{2}=\alpha\). Substituting \eqref{Sclosed} into the expression for \(\langle t_0 \rangle\) and simplifying the prefactors yields
\begin{equation}
\begin{aligned}
\langle t_0 \rangle
&= \frac{4r_0^2\alpha^2}{\pi D}\cdot\frac{1}{\pi^2}\,S\!\Big(\frac{\pi\theta_0}{\alpha},\frac{2\alpha}{\pi}\Big) \\[4pt]
&= \frac{4r_0^2\alpha^2}{\pi D}\cdot\frac{1}{\pi^2}\cdot
\frac{\pi}{4\Lambda^2}\Big(-1+\cos(2\theta_0)+\tan(\alpha)\sin(2\theta_0)\Big) \\[4pt]
&= \frac{r_0^2}{4D}\Big(-1+\cos(2\theta_0)+\tan(\alpha)\sin(2\theta_0)\Big).
\end{aligned}
\end{equation}

Therefore we arrive at a closed form expression for the mean first-passage time
\begin{equation}\label{closed-series-result}
\begin{split}
    \langle t_0 \rangle&=\frac{4r_0^2\alpha^2}{\pi D}\sum_{m=0}^{\infty}
\frac{\sin\!\big(\tfrac{(2m+1)\pi\theta_0}{\alpha}\big)}
{(2m+1)\big((2m+1)^2\pi^2-4\alpha^2\big)} \\
&= \frac{r_0^2}{2D} \sin (\theta_0 )\big(\tan (\alpha ) \cos (\theta_0 )-\sin (\theta_0 )\big)\\
&=\frac{r_ {0}^2}{4D}\left(\frac {\cos\left(\alpha[1-\frac{2\theta_0}{\alpha}] \right)}{\cos(\alpha)} - 1 \right).
\end{split}
\end{equation}
%The expression \eqref{closed-series-result} can also be presented more compactly as in \eqref{firstmomentclosedintext}, indicating the symmetry about the wedge center($\theta_0=\alpha/2$). 
An equivalent expression for a rotated infinite wedge with the wedge center aligned with positive $x-$axis (Absorbing boundaries at $\theta=\pm\alpha/2$) can be found in \cite{Redner_2001}.
\subsection{\textbf{Derivation of the Closed form of second moment \(\big\langle t_0^2\big\rangle\)}}\label{secmomentclosed}
\subsection*{}
\noindent Following the same methods in (\textbf{{\ref{firmomclosed}}}), a closed form for the following series can also be derived
\begin{equation}\label{secondmomentseries}
\big\langle t_0^2\big\rangle
= \frac{8 r_0^4 \alpha^4}{\pi D^2}\sum_{m=0}^{\infty}
\frac{\sin\!\big(\tfrac{(2m+1)\pi \theta_0}{\alpha}\big)}{(2m+1)}
\frac{1}{\big((2m+1)^2\pi^2 -16\alpha^2\big)\big((2m+1)^2\pi^2 -4\alpha^2\big)}.
\end{equation}

As before we assume the range and region of applicability,
\[
0<\theta_0<\alpha,\qquad \alpha\not\in\Big\{\frac{(2n+1)\pi}{4},\frac{(2n+1)\pi}{2}:n\in\mathbb Z\Big\},
\]
so that all denominators are nonzero.

Put \(k=2m+1\) (odd positive integers), and introduce
\[
x=\frac{\pi\theta_0}{\alpha},\qquad \Lambda=\frac{2\alpha}{\pi},
\]
so \(0<x<\pi\) and \(\Lambda>0\). The arguments in the denominator \eqref{secondmomentseries} therefore becomes,
\[
(2m+1)^2\pi^2-4\alpha^2 = \pi^2\big(k^2-\Lambda^2\big),\qquad
(2m+1)^2\pi^2-16\alpha^2 = \pi^2\big(k^2-4\Lambda^2\big).
\]
Hence the sum in \eqref{secondmomentseries} may be written as
\begin{equation}
\big\langle t_0^2\big\rangle
= \frac{8 r_0^4 \alpha^4}{\pi D^2}\cdot\frac{1}{\pi^4}
\sum_{\substack{k}}
\frac{\sin(kx)}{k\,(k^2-\Lambda^2)(k^2-4\Lambda^2)}.
\end{equation}
Let's define the sum
\begin{equation}\label{S2def}
S_2(x,\Lambda)=\sum_{\substack{k}}
\frac{\sin(kx)}{k\,(k^2-\Lambda^2)(k^2-4\Lambda^2)}.
\end{equation}
Hence the entire task now reduces to evaluating \(S_2(x,\Lambda)\).
\vspace{5pt}
\par
\noindent Inside the sum we perform partial fractions in \(k^2\):
\[
\frac{1}{(k^2-\Lambda^2)(k^2-4\Lambda^2)}
=\frac{1}{3\Lambda^2}\Big(\frac{1}{k^2-4\Lambda^2}-\frac{1}{k^2-\Lambda^2}\Big).
\]
which makes the sum $S_2(x,\Lambda)$ in its analytically tractable form
\begin{equation}
S_2(x,\Lambda)=\frac{1}{3\Lambda^2}\sum_{k}\left[
\frac{\sin(kx)}{k(k^2-4\Lambda^2)} - \frac{\sin(kx)}{k(k^2-\Lambda^2)}
\right].
\end{equation}
The above sum now contains two separate but identical terms, which we recognize from the previous derivation. Let's represent them as a common sum term
\[
S(x,\mu)=\sum_{\substack{k}}\frac{\sin(kx)}{k(k^2-\mu^2)},
\]
where \(\mu=2\Lambda\) or \(\mu=\Lambda\). Thus we can rewrite the series $S_2(x,\mu)$ as
\begin{equation}\label{S2_reduce}
S_2(x,\Lambda)=\frac{1}{3\Lambda^2}\Big(S(x,2\Lambda)-S(x,\Lambda)\Big).
\end{equation}

\par
\noindent The evaluation of $S(x,\mu)$ type sum was done in \eqref{Sclosed}, which we directly use here. Thus for \(0<x<\pi\)
\begin{equation}\label{subaxiliaryfn}
S(x,\mu)=\frac{\pi}{4\mu^2}\Big(-1+\cos(\mu x)+\tan\!\Big(\frac{\mu\pi}{2}\Big)\sin(\mu x)\Big).
\end{equation}
Apply this with \(\mu=\Lambda\) and \(\mu=2\Lambda\). Using the original definitions of $x\;\text{and}\;\Lambda$ substitutions, one may note that,
\[
\Lambda x = 2\theta_0,\qquad 2\Lambda x = 4\theta_0,\qquad
\frac{\Lambda\pi}{2}=\alpha,\qquad \frac{2\Lambda\pi}{2}=\Lambda\pi=2\alpha,
\]
Therefore substituting these into \eref{subaxiliaryfn} and thus evaluating \eref{S2_reduce}

\begin{equation}\label{subaxiliary2}
\begin{split}
S(x,2\Lambda)-S(x,\Lambda)
&= \frac{\pi}{16\Lambda^2}\Big(-1+\cos(4\theta_0)+\tan(2\alpha)\sin(4\theta_0)\Big)- \frac{\pi}{4\Lambda^2}\Big(-1+\cos(2\theta_0)+\tan(\alpha)\sin(2\theta_0)\Big) \\
&= \frac{\pi}{16\Lambda^2}\Big(3 + \cos(4\theta_0) - 4\cos(2\theta_0)
+ \tan(2\alpha)\sin(4\theta_0) - 4\tan(\alpha)\sin(2\theta_0)\Big).
\end{split}
\end{equation}

\noindent Combine \eqref{S2_reduce} with the prefactor in \eqref{secondmomentseries}
\begin{equation}
\big\langle t_0^2\big\rangle
= \frac{8 r_0^4 \alpha^4}{\pi D^2}\cdot\frac{1}{\pi^4}\;S_2(x,\Lambda)
= \frac{8 r_0^4 \alpha^4}{\pi^5 D^2}\cdot\frac{1}{3\Lambda^2}\Big(S(x,2\Lambda)-S(x,\Lambda)\Big)
\end{equation}
Substitute the expression for \(S(x,2\Lambda)-S(x,\Lambda)\) above and simplify using \(\Lambda^2=4\alpha^2/\pi^2\). After straightforward simplification all prefactors reduce to \(r_0^4/(96 D^2)\) also further reducing the trigonometric terms in the bracket, giving the compact final result below

\begin{equation}\label{t0sq_closed_alt}
\big\langle t_0^2\big\rangle
= \frac{r_0^4}{96 D^2}\Big(\frac{\cos (2 (\alpha -2 \theta_0 ))}{\cos (2 \alpha )} -\frac {8 \sin (\theta_0 )\sin (\alpha -\theta_0 )}{ \cos (\alpha )}-1\Big).
\end{equation}
\subsection{Coefficient of variation ($CV$)}
Obtaining $CV$ is now only a matter of substitutions and few simple trigonometric simplifications. Using the expressions for $\langle t_0\rangle$ and $\langle t_0^2 \rangle$, we substitute them in the following definition for $CV$
\begin{equation}
    CV=\sqrt{\frac{\langle t_0^2\rangle}{\langle t_0\rangle^2}-1}
\end{equation}
Thus we obtain
\begin{equation}
\begin{split}
     CV(\theta_0,\alpha)&=\sqrt{\frac{\Big(\frac{\cos (2 (\alpha -2 \theta_0 ))}{\cos (2 \alpha )} -\frac {8 \sin (\theta_0 )\sin (\alpha -\theta_0 )}{ \cos (\alpha )}-1\Big)}{24 \Big(\sin (\theta_0 )\big(\tan (\alpha ) \cos (\theta_0 )-\sin (\theta_0 )\big)\Big)^2}-1}%\\&=\sqrt{\frac{\cos (\alpha ) (\cos (2 (\alpha -\theta_0 ))-3 \cos (2 \alpha )+\cos (2 \theta_0 )+1)}{12 \cos (2 \alpha ) \sin (\theta_0 ) \sin (\alpha -\theta_0 )}-1} 
     \\ &=\sqrt{\frac{2 (1-5 \cos (2 \alpha )) \cos (\alpha(1 -\frac{2\theta_0}{\alpha}) )+5 \cos (\alpha )+3 \cos (3 \alpha )}{24 \cos (2 \alpha ) \sin (\theta_0 ) \sin (\alpha -\theta_0 )}} \\&=\sqrt{\frac{2 (1-5 \cos (2 \alpha )) \cos (\alpha(1 -\frac{2\theta_0}{\alpha}))+5 \cos (\alpha )+3 \cos (3 \alpha )}{12 \cos (2 \alpha )\Big(\cos(\alpha(1 -\frac{2\theta_0 }{\alpha}))- \cos(\alpha )\Big)}}
\end{split}
\end{equation}
Since $0\leq \theta_0\leq\alpha$, introducing the dimensionless parameter $\Theta_0=\frac{\theta_0}{\alpha},\;\text{where}\;0<\Theta_0<1$ is a mathematical convenience that ensures the behavior of $CV$ can be mapped for the entire range of $\theta_0$ without always restricting it within the different values of $\alpha$. This substitution reproduces the same expressions as represented in the main text.
\section{Solving for critical Wedge angle \((\alpha_c)\)}\label{solve-alphac}

\noindent To obtain $\alpha_{c}^{max}$ in \eqref{alphacritical2}, we solve the transcendental equation
\begin{equation}\label{trig-eq}
9\cos(3\alpha_{c}^{max})\;-\;22\cos(2\alpha_{c}^{max})\;+\;11\cos(\alpha_{c}^{max})\;+\;2 \;=\; 0,
\end{equation}
for \(\alpha_{c}^{max}\). The calculation proceeds by reducing the trigonometric expressions to a polynomial in \(c=\cos\alpha_{c}^{max}\), factoring that polynomial, and then inverting the cosine.

\vspace{10pt}
Let's reduce \eqref{trig-eq} into a polynomial form using the standard triple- and double-angle identities
\[
\cos(3\alpha)=4\cos^{3}\alpha-3\cos\alpha,\qquad
\cos(2\alpha)=2\cos^{2}\alpha-1,
\]
and set \(c=\cos\alpha_{c}^{max}\). Substituting these into \eqref{trig-eq} gives
\begin{align}\label{cubic-c}
\nonumber 9(4c^3-3c)-22(2c^2-1)+11c+2 &= 0 \\
9c^3 -11c^2 -4c +6 \;&=\; 0
\end{align}

The cubic \eqref{cubic-c} can be factorized as
\begin{equation}\label{cubic-factor}
9c^3 -11c^2 -4c +6 \;=\; (c-1)\,(9c^2-2c-6).
\end{equation}
Hence the three real roots for \(c\) are
\begin{equation}\label{c-roots-exact}
c \in \left\{\, 1,\; \frac{1+\sqrt{55}}{9},\; \frac{1-\sqrt{55}}{9}\,\right\}.
\end{equation}

Recall \(c=\cos\alpha_{c}^{max}\). Thus the corresponding solutions for \(\alpha_{c}^{max}\) are
\begin{equation}\label{alphac-exact}
\alpha_{c}^{max} \;=\; \pm\arccos(c) \;+\; 2\pi n,\qquad n\in\mathbb{Z},
\end{equation}
with \(c\) taking each value in \eqref{c-roots-exact}.
Since \(\alpha_{c}^{max}\) is constrained to the typical wedge-angle range \(0<\alpha\le\pi/4\), the only nondegenerate physically relevant solution is
\begin{equation}\label{alphac-physical}
\;\alpha_{c}^{max} \;=\; \arccos\!\Big(\frac{1+\sqrt{55}}{9}\Big)
\approx 0.36216159\ \text{rad}\; \approx 20.75^\circ\;. \;
\end{equation}
(The solution \(\alpha_{c}^{max}=0\) is degenerate and \(\alpha_{c}^{max}\approx2.3629\) lies outside \([0,\pi/4]\).)

\section{Numerical simulation methods: Algorithm for Poissonian reset implementation for Brownian particle inside the Wedge domain.}\label{num_sim}
In this section we briefly describe the steps to implement the numerical simulation of Brownian dynamics interrupted by stochastic resetting inside the wedge domain. The apex of the wedge is placed at the origin of a Cartesian coordinate system on the 2D plane, with the absorbing boundary \(\theta=0\) aligned with the positive \(x\)-axis and the other absorbing boundary \(\theta=\alpha\) an infinite ray emanating from the origin at angle \(\alpha\) with respect to the \(x\)-axis.

\subsection{The dynamics}
Diffusion inside the wedge is modeled as continuum Brownian motion. A particle at time \(t\) updates its Cartesian coordinates \((x(t),y(t))\) over a timestep \(\Delta t\) according to the Langevin dynamics
\begin{equation}
\begin{split}\label{langevindynamics}
    x (t+\Delta t) &= x(t) + \sqrt{2 D \Delta t}\,\xi_x (t),\\
	y (t+\Delta t) &= y(t) + \sqrt{2 D \Delta t}\,\xi_y (t),
\end{split}
\end{equation}
where \(\xi_x(t)\) and \(\xi_y(t)\) are independent standard normal variates. Particles are initialized at \((x(0),y(0))=(r_0\cos\theta_0,r_0\sin\theta_0)\). The \(P\) trajectories independently evolve according to \eqref{langevindynamics} until absorption at a wedge boundary or until a maximum time \(T_{\max}=N\Delta t\) (see below).

\subsection{Enumerating first-passage statistics}
The simulation involves an ensemble of \(P\) trajectories, each constrained to propagate utmost \(N\) time steps of duration \(\Delta t\) each. Particle positions are stored in Cartesian coordinates and the polar angle is computed via the function \(\theta=\arctan(y/x)\). 

\paragraph{Absorption detection.} \label{measure_fpt}
A trajectory is considered absorbed if the line segment joining successive positions \((x_n,y_n)\to(x_{n+1},y_{n+1})\) intersects either ray. We detect such intersections with a standard line–ray intersection test and, when an intersection occurs within the timestep, linearly interpolate the sub-step hitting time \(\tau\in[0,\Delta t)\) to obtain the first-passage time \(t_f=t_n+\tau\). This within-step interpolation substantially reduces discretizations bias compared with checking angles only at discrete times, which assumes $\tau=\Delta t$.
\vspace{0.5em}
\paragraph{Recorded observables.}
\begin{itemize}
  \item \textbf{Unconditioned MFPT:} Let $t_f^{i}=n_f\Delta t$ be the time of first absorption at either boundary for the $i^{th}$ trajectory, The mean unconditioned MFPT \(\langle t_f\rangle\) is the ensemble average over all absorbed trajectories $p\leq P$. We follow (\ref{measure_fpt}) as the condition of absorption. Say at time step $t_{f-1}$ the particle is at $(x_{f-1},y_{f-1})$, inside the wedge domain and in the next time step $t_{f-1}+\Delta t$ if it triggers the condition $\arctan(y_f/x_f)>\alpha$, then the first passage time is noted as $t_f=t_{f-1}+\tau$, where $\tau$ is the time after $t_f$, the Brownian particle exactly crosses the boundary at $\theta=0\;\text{or}\;\alpha$. For sufficiently small $\Delta t$, we assume $\tau \approx\Delta t$.
  %$\tau=\epsilon\Delta t$. Here $\epsilon$ is the ratio $\sqrt{\frac{(y^2_{f-1}+x^2_{f-1})}{(y^2_{f}+x^2_{f})}}$.
  
  \item \textbf{Conditioned MFPTs:} if a trajectory exits at boundary \(i\in\{0,\alpha\}\) we record \(t_f^{\,i}\). The conditioned means \(\langle t_f^{\,0}\rangle\) and \(\langle t_f^{\,\alpha}\rangle\) are ensemble averages over the subset of trajectories exiting via the respective boundary.
  \item \textbf{Splitting probabilities:} if \(p_f^{\,i}\) trajectories (out of \(P^\prime\) that exited through any of the boundaries within the total time-steps $N$) exit via boundary \(i\), the empirical splitting probability is \(\varepsilon_i = p_f^{\,i}/P^\prime\).
\end{itemize}

Trajectories that reach \(T_{\max}=N\Delta t\) without absorption are treated as \emph{censored}; we record the censoring fraction and ensure it is negligible for the results presented. Convergence of \(\langle t_f\rangle\), \(\langle t_f^{\,i}\rangle\), and \(\varepsilon_i\) was tested by (i) halving \(\Delta t\) and (ii) increasing the ensemble size \(P\). %and/or the number of independent replicates \(K\); reported values correspond to parameter choices for which differences were below the stated tolerance.

\subsection{Incorporating resetting}
Resetting is implemented as a Poisson process of rate \(\lambda\). In the discrete-time production runs used for the figures, at each timestep a uniform random number \(\delta\in[0,1]\) is drawn and a reset is triggered if \(\delta\leq\lambda\Delta t\). When a reset occurs the particle is %instantaneously 
placed at the reset location \((x_{\rm res},y_{\rm res})=(r_{\rm 0}\cos\theta_{\rm 0},r_{\rm 0}\sin\theta_{\rm 0})\) in the timestep. Between resets particles evolve via Eq.~\eqref{langevindynamics}.% using Euler--Maruyama.

We have also validated the same using another simulation technique that is event driven type, wherein we sampled inter-reset times from \(\operatorname{Exp}(\lambda)\). If a reset occurs inside a timestep, split the timestep at the sampled reset time, propagate Brownian increments over the sub-intervals, and place the reset at the exact sampled time. This scheme recovers the results obtained from the continuous-time resetting strategy.

\subsubsection*{Algorithm (For discrete-time reset scheme 
production runs)}
\begin{enumerate}
  \item Set parameters \(D,\alpha,r_0,\theta_0,\lambda,\Delta t,N,P\).
  \item For each trajectory \(p=1,\dots,P\): set \((x,y)=(r_0\cos\theta_0,r_0\sin\theta_0)\), \(t\rightarrow0\).
    \begin{enumerate}
      \item Draw \(\delta\sim\mathrm{Uniform}(0,1)\). If \(\delta\le\lambda\Delta t\), set \((x,y)\rightarrow(x_{\rm res},y_{\rm res})\) (instantaneous reset).
      \item If $\delta>\lambda\Delta t$, draw \(\xi_x,\xi_y\sim\mathcal N(0,1)\) and propose
      \[
      x'\leftarrow x+\sqrt{2D\Delta t}\,\xi_x,\qquad y'\leftarrow y+\sqrt{2D\Delta t}\,\xi_y.
      \]
      \item Check segment \((x,y)\to(x',y')\) for intersection with rays \(\theta=0\) and \(\theta=\alpha\). If intersection occurs at subtime \(\tau\in[0,\Delta t)\), record \(t_f=t+\tau\) and the exit boundary, then stop this trajectory. Otherwise set \((x,y)\rightarrow(x',y')\), \(t\rightarrow t+\Delta t\) and continue.
      \item If \(t\ge T_{\max}=N\Delta t\) without absorption, mark trajectory as censored ( or Take $t_f=T_{max}$ for very large $N\Delta t$) and stop.
    \end{enumerate}
  \item After all \(P\) trajectories complete, compute MFPTs, conditioned means and splitting probabilities. 
\end{enumerate}
\subsubsection*{Practical notes}
All production simulations used a Mersenne Twister RNG (e.g., \texttt{numpy.random.Generator} or equivalent). 
We also want to mention a few typical practical guidelines used in our runs:
\begin{itemize}
  \item Ensure \(\lambda\Delta t \ll 1\) for the discrete-time reset approximation to be accurate.
  \item Verify convergence by halving \(\Delta t\) and/or doubling \(P\); require reported MFPTs and splitting probabilities to change by less than the stated tolerance(Deviation from the theoretical value).
  \item For large wedge angles, particles initiated near the wedge center (\(\theta_0=\alpha/2\)) a larger \(\Delta t\) may be acceptable, whereas initial positions near the boundaries require smaller \(\Delta t\) to avoid missing boundary crossings. For narrow wedges, a reasonable $N\Delta t$ can record $t_f$ with sufficient precision. In all cases \(N\Delta t\) large enough to capture long tails is ensured.
\end{itemize}

\end{document}